# Submillimeter and Far-Infrared Polarimetric Observations of Magnetic Fields in Star-Forming Regions

Kate Pattle[1]* and Laura Fissel[2]

[1] National Tsing Hua University, Hsinchu, Taiwan, [2] National Radio Astronomy Observatory, Charlottesville, VI, United States

Observations of star-forming regions by the current and upcoming generation of submillimeter polarimeters will shed new light on the evolution of magnetic fields over the cloud-to-core size scales involved in the early stages of the star formation process. Recent wide-area and high-sensitivity polarization observations have drawn attention to the challenges of modeling magnetic field structure of star forming regions, due to variations in dust polarization properties in the interstellar medium. However, these observations also for the first time provide sufficient information to begin to break the degeneracy between polarization efficiency variations and depolarization due to magnetic field sub-beam structure, and thus to accurately infer magnetic field properties in the star-forming interstellar medium. In this article we discuss submillimeter and far-infrared polarization observations of star-forming regions made with single-dish instruments. We summarize past, present and forthcoming single-dish instrumentation, and discuss techniques which have been developed or proposed to interpret polarization observations, both in order to infer the morphology and strength of the magnetic field, and in order to determine the environments in which dust polarization observations reliably trace the magnetic field. We review recent polarimetric observations of molecular clouds, filaments, and starless and protostellar cores, and discuss how the application of the full range of modern analysis techniques to recent observations will advance our understanding of the role played by the magnetic field in the early stages of star formation.

## 1. INTRODUCTION

In this chapter we discuss single-dish polarimetric observations of star-forming regions made at far-infrared and submillimeter wavelengths. These observations make use of the tendency for asymmetric dust grains to align with their long axis perpendicular to the local magnetic field direction (Davis and Greenstein, 1951; Andersson et al., 2015). Measurement of linearly polarized thermal radiation from dust is a technique which is now coming into its own: the current and forthcoming generation of polarimeters are permitting wide-area surveys in polarized light across the far-infrared and submillimeter wavelength regime. Such surveys represent a significant improvement over previous observations, and open up the properties of magnetic fields in a variety of star-forming environments to statistically rigorous analysis.

Polarized dust emission is a key tool for understanding the role of magnetic fields in star-forming regions, being a direct measurement of magnetic field morphology on most size scales and at most densities, from the diffuse ISM to the highest densities found in gravitationally bound cores, at which the coupling between magnetic field orientation and grain alignment is thought to break down (e.g., Jones et al., 2015). Emission polarimetry can also provide indirect measurements of magnetic field strength, most commonly through the (Davis-)Chandrasekhar-Fermi method (Davis, 1951; Chandrasekhar and Fermi, 1953), and of the dynamical importance of magnetic fields to molecular clouds (e.g., Soler et al., 2013).

Interpretation of emission polarization observations requires care, due to the degenerate plane-of-sky polarization patterns produced by various combinations of three-dimensional magnetic field geometry and variable efficiency of alignment of dust grains with the magnetic field. Breaking such degeneracies in order to accurately interpret polarization observations often requires comparison to models. However, few simulations of magnetized star formation to date have produced synthetic observations with which comparisons can be made, in part due to the past paucity of observations. In this chapter we discuss comparison of data to models where such comparisons exist. For in-depth discussion of numerical simulations of star formation, we refer the reader to Teyssier and Commerçon (under review).

In this chapter we focus on observations made with single-dish instrumentation. For a discussion of recent advances in interferometric polarimetry, we refer the reader to Hull and Zhang (2019). Discussion in this chapter is restricted to observations of polarized continuum emission from dust grains aligned with their local magnetic field direction. Polarization induced by scattering from dust grains is discussed by Hull and Zhang (2019).

This chapter is structured as follows: in section 2 we discuss past, current and forthcoming instrumentation. In section 3 we discuss methods by which polarization observations are interpreted. In section 4 we discuss observations of magnetic fields on the scale of molecular clouds. In section 5, we discuss observations of magnetic fields in Bok globules; in section 6 we discuss observations of magnetic fields in filaments; and in section 7 we discuss observations of magnetic fields in starless, prestellar and protostellar cores. Section 8 summarizes this chapter.

## 2. AN OVERVIEW OF POLARIMETERS

Thermal emission from dense molecular clouds is typically a few percent polarized, making the detection of their polarized radiation challenging. Determining linear polarization requires measurement of differential power for light with different orientations of $\vec{E}$, and is typically characterized by the Stokes parameters $I$, $Q$, and $U$:

$$Q = I_0 - I_{90} \tag{1}$$
$$U = I_{45} - I_{135}, \tag{2}$$

where $I_x$ indicates the polarized component of total intensity $I$, with $\vec{E}$ parallel to the on-sky angle $x$. The polarization angle $\theta$ and fraction of the radiation that is polarized $p$ can be measured from the Stokes parameters with

$$\theta = \frac{1}{2}\arctan(U, Q) \tag{3}$$

and

$$p = \frac{\sqrt{Q^2 + U^2}}{I} \tag{4}$$

respectively. Here we have used the IAU convention that a polarization angle of $0°$ is aligned North-South and that $\theta$ increases when rotated toward the East of North.

In this section we discuss design and observation strategies of different types of polarimeters. We also briefly review polarimeters past, currently operating or being constructed, and proposed for next-generation far-IR/sub-mm satellites.

### 2.1. Polarimeter Design and Observation Strategies

Measurements of linear polarization with incoherent detectors, such as bolometers, require a method of measuring total power at different $E$-field orientations. Fast modulation of the polarized signal is also required so that $Q$ and $U$ can be measured on timescales faster than noise drifts associated with the instrument and/or observations. Finally, the background signal contributed from sky emission must be removed from the observations.

**Table 1** summarizes all single-dish polarimeters that have operated between 100 $\mu m$ to 1.2 mm and have resolution $< 10'$ FWHM, in addition to polarimeters that are being constructed, or have recently been proposed. The development of sub-mm and far-IR polarimeters has been driven by a quest for improvements in mapping speed, by increasing the number of detectors and operating at better observing sites. Ground-based polarimeters built for large-aperture telescopes such as the James Clerk Maxwell Telescope (JCMT, 15-m), Caltech Sub-mm Observatory (CSO, 10.4-m), Atacama Pathfinder EXperiment (APEX, 12-m), Institut de Radioastronomie Millimétrique (IRAM, 30-m), and the Large Millimeter Telescope (LMT, 50-m), can be used to make high-resolution maps of magnetic field morphology in star-forming regions. However, this resolution comes at the cost of observing through the atmosphere, requiring these polarimeters to observe through narrow windows in the sub-mm atmospheric transmission spectrum, or at millimeter bands away from the spectral peak of molecular cloud dust. The atmosphere also emits radiation at far-IR, sub-mm, and millimeter wavelengths, resulting in additional power absorbed by the detectors, or "loading," and reduces the overall detector responsivity.

Ideally, one would put sub-mm polarimeters in space (for example the *Planck* Surveyor); however, such satellites are

TABLE 1 | Listing of past, current, and proposed Sub-mm polarimeters.

| Instrument (Telescope) | Platform | Start date | Status | Bands [μm] | $N_{pixels}$ | FWHM [arcsec] | Polarization strategy | Key References |
|---|---|---|---|---|---|---|---|---|
| UCL Polarimeter | Balloon | 1980 | Finished | 77 | 1 | 300 | RP, Chopping | Cudlip et al., 1982 |
| KAO | Airborne | 1983 | Finished | 270 | 1 | 60 | S-HWP, K-mirror | Dragovan, 1986 |
| POLY (KAO) | Airborne | 1986 | Finished | 100 | 1 | 55, 40 | S-HWP | Novak et al., 1989 |
| MILLIPOL (NRAO-12m) | Ground | 1987 | Finished | 1300 | 1 | 30 | S-HWP, linear feed | Barvainis et al., 1988, Clemens et al., 1990 |
| UKT-Polarimeter (JCMT) | Ground | 1989 | Finished | 450, 850, 1100 | 1 | 8–18 | S-HWP, F-A | Flett and Murray, 1991 |
| Stokes (KAO) | Airborne | 1990 | Finished | 100 | 2 × 32 | 35 | S-HWP, P-G, Chop | Platt et al., 1991 |
| HERTZ (CSO) | Ground | 1995 | Finished | 353 | 2 × 32 | 20 | S-HWP, P-G, Chop | Schleuning et al., 1997, Dowell et al., 1998 |
| SCUPOL (JCMT) | Ground | 1997 | Finished | 850 | 37 | 14 | S-HWP, F-A, Chop | Murray et al., 1997, Greaves et al., 2003 |
| SPARO (VIPER) | Ground | 2000 | Finished | 450 | 2 × 9 | ~300 | S-HWP, P-G, Chop | Dotson et al., 1998, Renbarger et al., 2004 |
| BLASTPol | Balloon | 2010 | Finished | 250, 350, 500 | 266 | 150 | S-HWP, P-G, Scan | Galitzki et al., 2014b |
| SHARP/SHARC (CSO) | Ground | 2005 | Finished | 350 and 450 | | 9 | X-Grid, S-HWP, chopping | Li et al., 2008 |
| PolKa/LABOCA (APEX) | Ground | 2011 | Not Currently Offered | 870 | 295 | 20 | RPM | Siringo et al., 2012, Wiesemeyer et al., 2014 |
| HAWC+ (SOFIA) | Airborne | 2016 | Active | 53, 62, 89, 154, 214 | 1,280 | 4.8–18.2 | S-HWP, Dual-Pol, Chop | Dowell et al., 2018 |
| Planck | Space | 2009 | Finished | 850 | 8 | 300 | Scan | Lamarre et al., 2010, Planck Collaboration VIII, 2016 |
| PILOT | Balloon | 2015 | Last Flight 2017 | 214 | 2048 | 120 | S-HWP, dual-analyzer, Scan | Foënard et al., 2018 |
| POL2 (JCMT) | Ground | 2016 | Active | 450, 850 | 5120 (each) | 10, 14 | Sp-HWP, Dual-Pol | Bastien et al., 2011, Friberg et al., 2016 |
| BLAST-TNG | Balloon | 2019 | Integration | 250, 350, 500 | 759, 475, 230 | 31, 41, 59 | S-HWP, Dual-Pol, Scan | Galitzki et al., 2014a |
| TolTEC (LMT) | Ground | 2019 | Construction | 1100, 1400, 2100 | 900, 1800, 3600 | 5.0, 6.3, 9.8 | Sp-HWP, Dual-Pol | Bryan et al., 2018 |
| NIKA-2 (IRAM) | Ground | 2019 | Integration | 1150, 2000 | 1140 × 2616 | 11, 18 | Dual-Pol, HWP | Adam et al., 2018 |
| A-MKID (APEX) | Ground | 2019 | Integration | 350, 850 | 3520 (350), 21600 (850) | 19 (at 850) | Dual-Pol, Sp-HWP | Otal, 2014 |
| POL (SPICA) | Space | ~2030 | Proposed | 100, 200, 350 | 32 × 32, 16 × 16, 8 × 8 | 9, 18, 32 | S-HWP, Dual-Pol, Scan | Gaspar Venancio et al., 2017, Roelfsema et al., 2018 |
| PICO | Space | ~2030 | Proposed | 375, 450, 541, 649, 779, 935, 1124 | 10000 total | 66–192 | Dual-Pol, Scan | Sutin et al., 2018, Young et al., 2018 |
| FIP (OST) | Space | ~2030 | Proposed | 50, 250 | 26000/6500 | 2–10 | Dual-Pol, Sp-HWP, Scan | Staguhn et al., 2018 (Concept-2) |

Polarization Strategy Abbreviations: S-HWP, Stepped HWP; Sp-HWP, Continuously Spinning HWP; P-G, Polarizing Grid; X-Grid, Cross-Grid Splitter; RP, Rotating Polarizing Grid; RPM, Reflecting Polarization Modulator; F-A, Fixed Analyzer/Polarizing Grid; Dual-Pol, Dual Polarization Detectors; Scan, Scan Modulation; Chop, Chopping.

very expensive. Alternatively, polarimeters can operate in the stratosphere. Polarimeters on an aircraft, such as the Kuiper Airborne Observatory (KAO) or Stratospheric Observatory for Infrared Astronomy (SOFIA), typically operate at 10–13 km above sea-level (∼90 % of the atmosphere), which greatly decreases the atmospheric loading and allows observations at wavelengths $< 300\,\mu m$. Stratospheric balloons offer an even more lofty platform at 35–50 km above sea-level (above $> 99\%$ of the atmosphere). Such balloon-borne polarimeters can operate in near-space conditions, at a fraction of the cost of a satellite. However, stratospheric balloon flights are currently limited to several weeks' length, reducing the amount of polarization data obtainable.

### 2.1.1. Measuring Differential Power

The most basic requirement of a polarimeter is to measure the intensity of the component of incoming radiation at different polarization angles. This has often been accomplished by placing a polarizing grid in the light path detector focal plane, such that component of the radiation with $\vec{E}$-orientation parallel to the grid wires is reflected while radiation with $\vec{E}$ perpendicular to the wires is transmitted. The reflected orthogonal polarization component can be directed to different focal plane arrays. This method is somewhat inefficient: either half the light is discarded before reaching the detector focal plane (e.g., BLASTPol), or one part of the array is used to detect one polarization component, and a separate part of the array is required to detect the orthogonal component (e.g., SHARP, SPARO polarimeters).

Most modern polarimeters now use dual-polarization detectors, which can measure both orthogonal polarization components at the same location on the focal plane, for example Transition Edge-Sensors (TES), and Kinetic Inductance Detectors (KIDs) (see Mauskopf, 2018 for a recent review).

### 2.1.2. Polarization Modulation

For ground-based polarimeters, the dominant source of noise comes from short-timescale fluctuations in the thermal emission of atmosphere and telescope. High-frequency referencing to an off-source sky position or fast ($>> 1$ Hz) modulation of the polarization orientation measured by the instrument is therefore crucial.

#### 2.1.2.1. Chopping

The noise of a polarization measurement can be reduced by high-frequency "chopping" of an optical element, commonly the secondary mirror, to a nearby location on the sky assumed to be free of polarized emission (see Hildebrand et al., 2000). The size of the pointing offset, or chop-throw, severely limits the largest angular scales that can typically be recovered. Also, if there is polarized emission at the reference locations then this will add a systematic error to the polarization measured at the target location (see Appendix A of Matthews et al., 2001b).

#### 2.1.2.2. Rotation of a half-wave plate

Birefringent half-wave plates (HWP) rotate the polarization angle of incoming light by $2\alpha$, where $\alpha$ is the angle of HWP rotation. These HWPs can serve two purposes: if spun continuously they can modulate the polarization such that all Stokes parameters can be measured on timescales faster than the low-frequency drifts of the telescope (e.g., POL2, TolTEC). In contrast, a stepped HWP can be used to rotate the polarization, in order to measure both Stokes $Q$ and $U$ with each individual detector, and thereby correct for differences in detector beam shape or gains, and characterize the instrumental polarization (IP).

HWPs are an important tool for modulating polarization. However, their disadvantages include modulation of polarization from the optical path between the source and the HWP, preventing their use to characterize IP caused by optical elements earlier in the light path, such as the primary and secondary mirrors. Also, any differences in transmission across the HWP can cause the signal incident to the detectors to vary. It is thus advisable to place a HWP before the re-imaging optics, and far from a focus point of the instrument.

#### 2.1.2.3. Modulation by scanning

For instruments where the time-scale associated with low-frequency ($1/f$) noise is long, polarization can be modulated by scanning the telescope such that Stokes $Q$ and $U$ can be measured at a given location on the sky on timescales faster than $(1/f)$.

This was the strategy adopted by the BLASTPol balloon-borne polarimeter (Galitzki et al., 2014b), which utilized a patterned polarization grid such that each adjacent bolometer sampled an orthogonal polarization component. As the telescope scanned across a target region, the time between when a source was measured with one detector and a detector sampling an orthogonal polarization component was $\ll 1$ s. The largest recoverable scale was therefore bounded by the scan speed/$(1/f)$. For BLASTPol a typical scan speed was $0.2°\,s^{-1}$, and the characteristic 1/f knee frequency was $\sim 50$ mHz, so polarized emission on the scales of several degrees could be recovered.

## 2.2. Previous Polarimeters
### 2.2.1. Early Detections of Polarized Emission

The first successful observation of linearly-polarized emission was by Cudlip et al. (1982), using the UCL 60 cm telescope, which operated from a stratospheric balloon platform, and used a fast-rotating polarized grid (32 Hz), combined with telescope chopping at 4 Hz. Cudlip et al. (1982) found a polarization level of 2.2% for the Orion Nebula integrated over a frequency band with an effective central wavelength of $77\,\mu m$ for a 70 K blackbody spectral shape, and measured a polarization angle that was roughly orthogonal to the polarization angle measured from the polarization of extincted starlight, suggesting that the polarized signal was indeed due to emission from dust grains aligned with long axes perpendicular to the magnetic field.

Later Hildebrand et al. (1984) made the first detection of sub-mm polarization centered at $270\,\mu m$, also of Orion-KL, using a $^3$He-cooled bolometer system on the KAO. They detected $1.7 \pm 0.4\,\%$ polarization, with a polarization orientation that agreed with the angle from Cudlip et al. (1982), using two different methods of modulating polarization: a rotating sapphire HWP and a rotating K-mirror, and found consistent polarization levels. This polarimeter was later reconstructed to operate at $100\,\mu m$, closer to the spectral peak of hot dust in bright active star-forming regions (Novak et al., 1989).

The first ground-based detection was made with the 1.3 mm MILLIPOL instrument on the NRAO 12-m telescope (Barvainis et al., 1988). MILLIPOL used a HWP rotating at 1.6 Hz to modulate polarization signal directed to a linearly-polarized feed, and again observed Orion-KL, finding polarization angles consistent with previous far-IR and sub-mm stratospheric observations. A polarimeter was also constructed for the UKT-14 single bolometer instrument on the JCMT (Flett and Murray, 1991).

### 2.2.2. Improvements in Polarimeters, 1990–2017

In the 1990s use of low-noise amplifiers, combined with the ability to construct large focal plane arrays of bolometers, made observations of larger areas and fainter sources possible. The first polarimeter using an array of bolometers was the STOKES instrument, which was built for the KAO. STOKES began operations in 1991 and had two arrays of 32 bolometers that simultaneously measured orthogonal polarization components (Platt et al., 1991). STOKES made over 1,100 individual polarization measurements during its 5 year operational period (Dotson et al., 2000).

Ground-based polarimeters also took advantage of sub-mm bolometer arrays, such as the Hertz instrument for the CSO (Schleuning et al., 1997; Dowell et al., 1998), and SCUPOL, built for the SCUBA camera at the JCMT (Murray et al., 1997; Greaves et al., 2003). Over a decade these two polarimeters observed dense sub-regions within molecular clouds, protostars, supernova remnants and bright nearby galaxies (see Matthews et al., 2009; Dotson et al., 2010 for summaries of the observations).

However, the necessity of instrument chopping to remove atmospheric noise made recovering polarization on scales larger than a few arcminutes difficult. The SPARO instrument, which operated on the 2-m VIPER telescope at the Amundsen-Scott South Pole station, took advantage of the atmospheric stability from the extremely cold and dry conditions during the Antarctic winter to use a much larger chop throw of $0.5°$, and therefore make the first large-scale polarization maps across entire molecular clouds (Li et al., 2006).

Later polarimeters were built with even larger detectors arrays, such as SHARP, which used a cross-grid to direct horizontally and vertically polarized light to opposite sides of the SHARC-II camera on the CSO, such that a $12 \times 12$ bolometer array would measure each polarization component (Li et al., 2008). The PolKa instrument on the Atacama Pathfinder Experiment (APEX, Güsten et al., 2006), was built for the LABOCA instrument, operating at 870 $\mu$m with 295 pixels (Wiesemeyer et al., 2014).

Sub-mm polarimetry from sub-orbital platforms on stratospheric balloon-borne telescopes also saw major advances. BLASTPol (the Balloon-borne Large Aperture Sub-mm Telescope for Polarimetry), simultaneously imaged the sky in three wide frequency ($\Delta f/f = 0.3$) passbands centered at 250, 350, and 500 $\mu$m (Galitzki et al., 2014b). During Antarctic science flights in 2010 and 2012 BLASTPol was been able to recover polarized emission on degree-scales, impossible for ground-based telescopes. The PILOT balloon-borne polarimeter, which operates at 214 $\mu$m and has even more detectors than BLASTPol, has flown from both Canada and Australia (Foënard et al., 2018).

Finally the *Planck* Surveyor, launched in 2007, was the first satellite polarimeter to both provide all-sky observations in the sub-mm (at 850 $\mu$m) and to have sufficient resolution to make fairly detailed (FWHM $\sim 5'$) maps of molecular clouds (Lamarre et al., 2010; Planck Collaboration VIII, 2016).

### 2.3. Current Instrumentation

Current polarimeters benefit from new technology which allows for the automated construction of large focal-plane arrays of detectors, such as the super-conducting transition-edge sensor (TES) bolometers, or kinetic inductance detectors (KIDs). In **Figure 1**, we compare the spatial-scale and instrument sensitivity to cold dust (left panel) and warm dust (right panel) for several recent, upcoming, and proposed polarimeters.

An exciting new instrument is the POL-2 polarimeter, which operates simultaneously at 450 and 850 $\mu$m and uses a half wave plate spinning at 2 Hz to measure linear polarization with the 10,000 pixels SCUBA-2 camera on the JCMT (Bastien et al., 2005; Friberg et al., 2016).

Additionally, the HAWC+ instrument on SOFIA has recently begun science operations (Harper et al., 2018). With 1280 TES bolometers and a best resolution of $5''$ at 53 $\mu$m HAWC$^+$ is producing high resolution maps of protostars and active star-forming regions.

### 2.4. Future Polarimeters

Several new polarimeters will be coming online in the next few years. One is the TolTEC camera on the newly-upgraded 50-meter Large Millimeter Telescope (LMT) in Puebla, Mexico, which should begin commissioning in early 2019 (Bryan et al., 2018). TolTEC uses microwave kinetic inductance detectors (mKIDs, Austermann et al., 2018) and operates simultaneously at 1.1, 1.4, and 2.1 mm. With $5''$ FWHM resolution at 1.1 mm TolTEC will have a factor of two improvement in resolution compared to any other single-dish sub-mm or millimeter polarimeter. Commissioning is also underway for the NIKA-2 (Adam et al., 2018) and A-KIDs (Otal, 2014) mKID array mm/sub-mm cameras on the IRAM/APEX telescopes, with instruments expected to include polarimetry capability. These new higher-mapping-speed, high-resolution instruments will be extremely important for mapping magnetic fields within filaments and dense cores.

High-detail maps of magnetic fields covering entire molecular clouds are the goal of the next-generation BLAST telescope (BLAST-TNG; Galitzki et al., 2014a). BLAST-TNG is expected to launch in December 2019 from McMurdo Station, Antarctica for a $\sim$28 day flight, and will map dozens of molecular clouds. The new version of BLAST-TNG also uses large-format mKID arrays, with an expected >10 times increase in mapping speed and $\sim$5 times increase in resolution compared to BLASTPol.

Finally, several satellite telescopes have been proposed that include far-IR, sub-mm, and mm linear polarization sensitivity. These telescopes would be cooled to $\leq 6$ K, and consequently the instrumental loading would be much lower than that of ground-based or stratospheric polarimeters. The design for the SPICA satellite currently includes a sub-mm polarimeter which would operate at 100, 200, and 350 $\mu$m (Gaspar Venancio et al., 2017; Roelfsema et al., 2018), while the Concept-2 design for the proposed Origins Space Telescope includes a Far-infrared Polarimeter (FIP), which would operate in both the far-IR and sub-mm (Staguhn et al., 2018). Both satellites would map hundreds of square degrees, with $\sim$10″, at 100 and 250 $\mu$m, respectively. A satellite targeting the entire sky, the Probe for Inflation and Cosmic Origins (PICO), has also been proposed (Sutin et al., 2018; Young et al., 2018). PICO would provide coverage in 10 frequency bands between 375 $\mu$m to 2 mm, with a best resolution of 1.1′. While this resolution is considerably lower than that of SPICA and OST, PICO would map every molecular cloud in the Galaxy, with thousands of molecular clouds mapped to a resolution better than 1 pc.

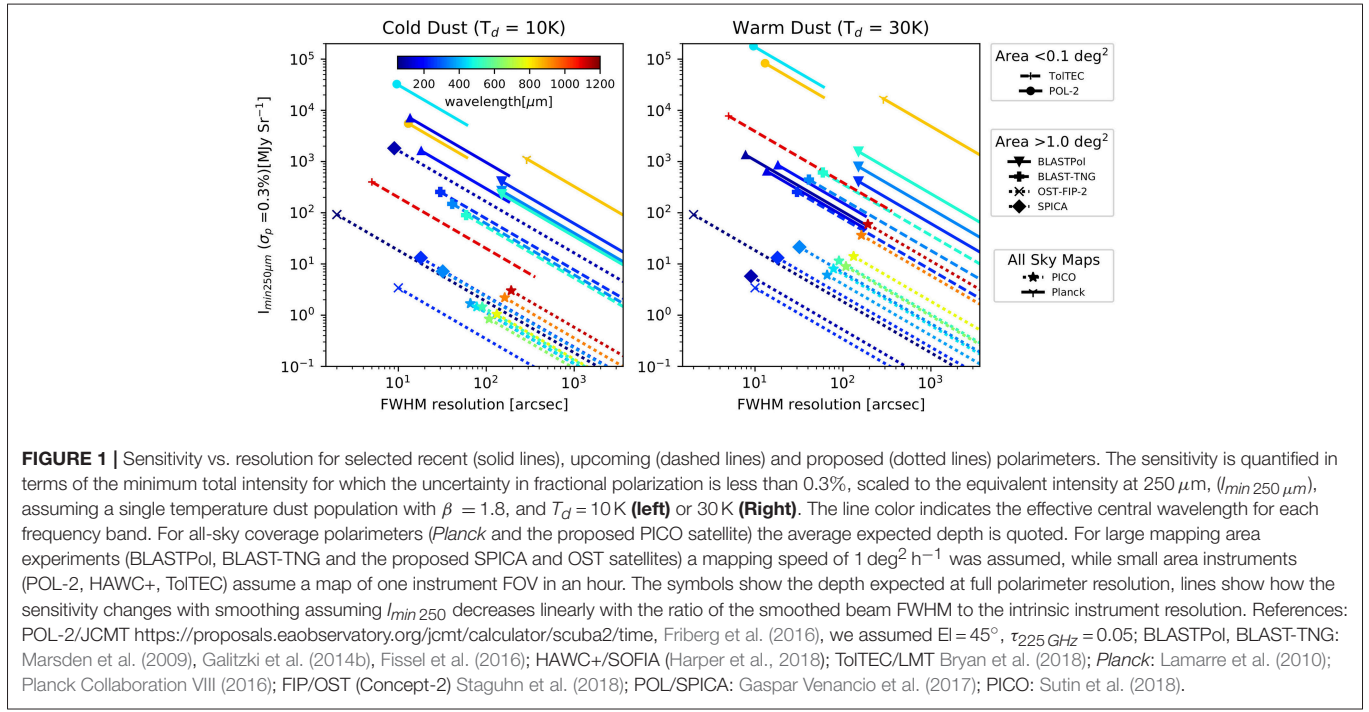

**FIGURE 1** | Sensitivity vs. resolution for selected recent (solid lines), upcoming (dashed lines) and proposed (dotted lines) polarimeters. The sensitivity is quantified in terms of the minimum total intensity for which the uncertainty in fractional polarization is less than 0.3%, scaled to the equivalent intensity at 250 $\mu$m, ($I_{min\,250\,\mu m}$), assuming a single temperature dust population with $\beta$ = 1.8, and $T_d$ = 10 K **(left)** or 30 K **(Right)**. The line color indicates the effective central wavelength for each frequency band. For all-sky coverage polarimeters (*Planck* and the proposed PICO satellite) the average expected depth is quoted. For large mapping area experiments (BLASTPol, BLAST-TNG and the proposed SPICA and OST satellites) a mapping speed of 1 deg$^2$ h$^{-1}$ was assumed, while small area instruments (POL-2, HAWC+, TolTEC) assume a map of one instrument FOV in an hour. The symbols show the depth expected at full polarimeter resolution, lines show how the sensitivity changes with smoothing assuming $I_{min\,250}$ decreases linearly with the ratio of the smoothed beam FWHM to the intrinsic instrument resolution. References: POL-2/JCMT https://proposals.eaobservatory.org/jcmt/calculator/scuba2/time, Friberg et al. (2016), we assumed El = 45°, $\tau_{225\,GHz}$ = 0.05; BLASTPol, BLAST-TNG: Marsden et al. (2009), Galitzki et al. (2014b), Fissel et al. (2016); HAWC+/SOFIA (Harper et al., 2018); TolTEC/LMT Bryan et al. (2018); *Planck*: Lamarre et al. (2010); Planck Collaboration VIII (2016); FIP/OST (Concept-2) Staguhn et al. (2018); POL/SPICA: Gaspar Venancio et al. (2017); PICO: Sutin et al. (2018).

## 3. TECHNIQUES FOR INTERPRETATION OF POLARIZATION OBSERVATIONS

In this section we summarize techniques for interpreting polarization observations, particularly in terms of determining the strength and energetic importance of the magnetic field. We discuss degeneracies between grain alignment, line-of-sight and sub-beam effects which complicate determination of magnetic field properties from polarization observations.

## 3.1. (Davis-)Chandrasekhar-Fermi Analysis and Its Variations

The most widely-used method of estimating magnetic field strength from continuum polarization data is the (Davis-)Chandrasekhar-Fermi (DCF) method (Davis, 1951; Chandrasekhar and Fermi, 1953). DCF assumes perturbations in the magnetic field to be Alfvénic, i.e., deviation in angle from the mean field direction results from distortion by small-scale non-thermal motions, such that $\delta v \propto \delta B/\sqrt{\rho}$. The plane-of-sky magnetic field strength is estimated as

$$B_{pos} = Q\sqrt{4\pi\rho}\frac{\sigma_v}{\sigma_\theta}, \qquad (5)$$

where $\rho$ is volume density, $\sigma_v$ is velocity dispersion, $\sigma_\theta$ is dispersion in angle, and $Q$ is a correction factor discussed below. DCF further assumes that turbulence is statistically isotropic, i.e., $\sigma_{v,los} = \sigma_{v,pos}$ (LOS – line of sight; POS – plane of sky).

Numerous attempts at improving the DCF method exist, falling into two camps: (1) better estimation of $\sigma_\theta$ and $Q$, (2) direct measurement of the ratio of turbulent to ordered magnetic energy through structure function analyses.

### 3.1.1. Classical DCF Method

Classical DCF assumes that the turbulent-to-ordered magnetic field strength ratio $\frac{B_t}{B_o} \sim \sigma_\theta/Q$, and that variation about the mean field direction is Gaussian and results from turbulent fluctuations about the mean field direction (the effect of measurement uncertainty on the measured dispersion in angle can where necessary be accounted for; see Pattle et al., 2017).

*3.1.1.1. Q parameter*
Classical DCF overestimates magnetic field strength due to two integration effects, (1) of ordered structure on scales smaller than the telescope beam, and (2) of emission from multiple turbulent cells within the telescope beam, including those along the line of sight. These effects are parameterized as a correction factor, $0 < Q < 1$ (Zweibel, 1990; Myers and Goodman, 1991; Ostriker et al., 2001).

Heitsch et al. (2001) found, based on numerical simulations, that for strong magnetic fields with well-resolved field structure, DCF results are typically correct to within a factor of 2, but strengths of weak and/or poorly-resolved fields could be overestimated by a factor $\lesssim 10$. Padoan et al. (2001) found that in low- to intermediate-density regions, $Q \sim 0.3 - 0.4$, while Ostriker et al. (2001) found $Q \sim 0.5$ at high densities. Crutcher et al. (2004) thus suggested that in dense, self-gravitating filaments and cores in which little field substructure is expected, $Q \sim 0.5$ is a reasonable value.

DCF will overestimate field strength by a factor $\sqrt{N}$, where $N$ is the number of turbulent cells enclosed within the volume sampled by the telescope beam (e.g., Houde et al., 2009). Where the linear resolution of the telescope beam is smaller than the scale of a turbulent cell, this overestimation reduces to the square root of the number of turbulent cells along the line of sight: $\sim \sqrt{L_{los}/L_{turb}} \sim \sqrt{N}$, where $L_{los}$ is the length of the optically thin column along the line of sight and $L_{turb}$ is the driving length scale of the turbulence (e.g., Cho and Yoo, 2016). Cho and Yoo (2016) proposed a measure of $N$, and so correction specifically for line-of-sight variations,

$$\frac{\delta V_c}{\delta v_{los}} \sim \frac{1}{\sqrt{N}}, \tag{6}$$

where $\delta V_c$ is the standard deviation of centroid velocities across the area to which the DCF method is applied, and $\delta v_{los}$ is the average line-of-sight velocity dispersion across the same region. While a correction for sub-beam effects is also necessary, an independent estimate of $N$ provides a useful check on other methods of parameterizing line-of-sight effects, as described below.

Interferometric results show complex magnetic field structure on small scales within molecular clouds (e.g., Hull et al., 2017), suggesting that DCF analyses using single-dish data need a good understanding of the effect of sub-beam field structure—particularly, ordered structure with size scales potentially smaller than the turbulent dissipation scale of the system—on measured angular dispersion.

*3.1.1.2. Large-scale ordered field structure*
DCF assumes that all variation in the magnetic field direction results from perturbations driven by Alfvénic turbulence, i.e., that the underlying field geometry is linear, which is not generally the case.

Pillai et al. (2015) introduced a "spatial filtering" method to account for ordered field structure, in which at each position the ordered field component is approximated by a distance-weighted mean of the angle at neighboring positions. The residual angle is given by

$$\theta_{i,res} = \theta_i - \frac{\sum_{j=1}^{N} w_{i,j}\theta_j}{\sum_{j=1}^{N} w_{i,j}}, \tag{7}$$

where the weighting function $w_{i,j} = \sqrt{1/d_{i,j}}$, and $d_{i,j}$ is the separation between positions $i$ and $j$. The angular dispersion $\sigma_\theta$ is determined from the standard deviation of these residuals.

Similarly, Pattle et al. (2017) introduced an "unsharp masking" method, in which the map of magnetic field angles is smoothed with a $3 \times 3$-pixel boxcar filter. This smoothed map—a model of the underlying ordered field—is subtracted from the original map, and the angular dispersion $\sigma_\theta$ is determined from the standard deviation of the residuals.

These methods require a separate estimate of the $Q$ parameter.

*3.1.1.3. Restrictions on angular dispersion*
Classical DCF is valid in the small-angle limit, found to be $\sigma_\theta \lesssim 25°$ (Ostriker et al., 2001; Padoan et al., 2001). Heitsch et al. (2001) present a correction for the small-angle approximation, $\sigma_\theta \rightarrow \sigma(\tan\theta)$, in Equation (5); this requires further correction in order to avoid anomalous behavior as $\theta \rightarrow \pm 90°$. Falceta-Gonçalves et al. (2008) present a more generalized DCF equation,

$$B_{pos}^{ext} + \delta B \simeq \sqrt{4\pi\rho}\frac{\sigma_v}{\tan\sigma_\theta}, \tag{8}$$

where $B_{pos}^{ext}$ is the plane-of-sky projected component of the mean (ordered) magnetic field and $\delta B$ is the turbulent field component, taking $\sigma(\tan\theta) \approx \tan\sigma_\theta$ to avoid discontinuities.

### 3.1.2. Structure-Function DCF Method
An alternative approach is to invoke structure function analysis to determine the ratio of the turbulent to the total magnetic field strength. This was first applied to the DCF method by Falceta-Gonçalves et al. (2008) and expanded upon by Hildebrand et al. (2009) (accounting for large-scale field structure), and Houde et al. (2009) (additionally accounting for sub-beam and line-of-sight effects).

In structure function analyses, the DCF equation is modified to become

$$B_{pos} = \sqrt{4\pi\rho}\sigma_v \left(\frac{\langle B_t^2 \rangle}{\langle B_o^2 \rangle}\right)^{-\frac{1}{2}} \tag{9}$$

where $B_t$ is the turbulent component of the magnetic field and $B_o$ is the ordered component of the magnetic field, such that

$$\langle B^2 \rangle = \langle B_t^2 \rangle + \langle B_o^2 \rangle, \tag{10}$$

with $B$ being total magnetic field strength.

The structure function under consideration is the average difference in angle between pairs of measured polarization vectors at positions $\vec{r}$ and $\vec{r} + l$ as a function of the distance $l$ between them,

$$\langle \Delta\theta(l) \rangle = \langle \theta(\vec{r}) - \theta(\vec{r} + \vec{l}) \rangle. \tag{11}$$

Houde et al. (2009) fit the function

$$1 - \langle \cos[\Delta\theta(l)] \rangle \simeq \frac{1}{N(\delta, W, \Delta')} \frac{\langle B_t^2 \rangle}{\langle B_o^2 \rangle}\left(1 - \exp\left[\frac{l^2}{2(\delta^2 + 2W^2)}\right]\right) + al^2 \tag{12}$$

where $N$ is given by

$$N(\delta, W, \Delta') = \Delta'\frac{\delta^2 + 2W^2}{\delta^3\sqrt{2\pi}}. \tag{13}$$

See Hildebrand et al. (2009) and Houde et al. (2009) for the derivation of this result. This function is fitted for $\langle B_t^2\rangle/\langle B_o^2\rangle$, the mean ratio of the turbulent and ordered field components; $\delta$, the turbulent length scale; and $a$, the first term in the Taylor expansion of the autocorrelation function. Fixed quantities are $W$, the telescope beam width ($FWHM = W\sqrt{8\ln 2}$), and $\Delta'$, the effective cloud thickness, which is assumed by Houde et al. (2009) to be the FWHM of the autocorrelation function of the polarized flux emission as a function of distance $l$.

This method requires the turbulent length scale $\delta$ to be resolved by the observations in order to determine $N$ and $\langle B_t^2\rangle/\langle B_o^2\rangle$. Where $\delta$ is not resolved ($\delta \lesssim W$), the maximum value of $N$ can be constrained, for an assumed cloud thickness $\Delta'$ (Pillai et al., 2015).

### 3.1.3. Correction for Total Field Strength

Total magnetic field strength can be estimated by combining DCF plane-of-sky measurements with line-of-sight field strengths determined from Zeeman splitting (e.g., Kirk et al., 2006), requiring the Zeeman-split line emission and the polarized dust emission to trace the same material. Kirk et al. (2006) combine JCMT/SCUPOL-data-derived DCF estimates with Zeeman splitting of the high-density tracer CN to estimate total field strength in a dense core. However, as Zeeman splitting is most easily detected in HI and OH (e.g., Crutcher, 2012), this analysis is more easily applicable at low-to-intermediate densities. Comparison of DCF and Zeeman measurements is discussed in detail in section 3.2.

The total magnetic field strength can alternatively be estimated statistically. Crutcher et al. (2004) argue that for a magnetic field geometry without a preferred axis, on average,

$$B_{pos} = \frac{\pi}{4}|\vec{B}|, \qquad (14)$$

where $|\vec{B}|$ is the magnitude of the total magnetic field strength. While this correction is useful when considering an ensemble of DCF measurements (e.g., Crutcher et al., 2004), its applicability to any individual case is uncertain.

Heitsch et al. (2001) proposed a statistical correction for line-of-sight effects, where the full magnetic field strength is estimated as

$$|\vec{B}| = \left[4\pi\rho\frac{\sigma_v}{\sigma(\tan\theta)}\left(1 + 3\sigma(\tan\theta)^2\right)\right]^{0.5}. \qquad (15)$$

### 3.1.4. Comparison and Testing of DCF Methods

Only classical DCF has been fully tested against synthetic observations generated from MHD simulations including self-gravity (Heitsch et al., 2001; Ostriker et al., 2001; Padoan et al., 2001), thus determining $Q \sim 0.5$. The principle of the structure function method has been tested against non-self-gravitating simulations (Falceta-Gonçalves et al., 2008), but its practical realizations (Hildebrand et al., 2009; Houde et al., 2009) have not been. Additionally, Pattle et al. (2017) performed limited testing of their "unsharp masking" variant of the classical DCF method against model field geometries.

Some direct comparisons have been made of the various DCF methods. Poidevin et al. (2013), using JCMT/SCUPOL data, compared the Falceta-Gonçalves et al. (2008) classical modification to the Hildebrand et al. (2009) structure function method, finding Hildebrand et al. (2009) field strengths to be consistently higher, typically by a factor $\sim 5$, across a range of densities ($\sim 10^3$—$10^6$ cm$^{-3}$), a difference ascribed both to high angular dispersions affecting the Falceta-Gonçalves et al. (2008) estimates and to the lack of correction for line-of-sight or beam signal integration effects in the Hildebrand et al. (2009) method. Planck Collaboration Int. XXXV (2016) compared the classical DCF method to the Houde et al. (2009) structure function method, finding that at low densities ($n \sim 100$ cm$^{-3}$) and high angular dispersions ($\sigma_\theta > 25°$), the Houde et al. (2009) method gives field strengths approximately twice those of classical DCF.

Hildebrand et al. (2009) and Pattle et al. (2017) found comparable values for magnetic field strength in OMC 1, of 3.8 mG (CSO/Hertz data) and $6.6 \pm 4.7$ mG (JCMT/POL-2 data), respectively. Houde et al. (2009) found 0.76 mG (CSO/SHARP) for the same region, inferring $N \sim 21$ independent turbulent cells along the line of sight. Using the Cho and Yoo (2016) method and C$^{18}$O line data, Pattle et al. (2017) found $N \lesssim 2$.

A self-consistent comparison of the various classical and structure-function DCF methods, using both observational data and synthetic observations from self-gravitating MHD simulations, would be of significant value. Comparison of the values of $N$ determined by the structure-function method and by the Cho and Yoo (2016) parameterization would also be useful.

In **Figure 2** we collate DCF-estimated magnetic field strengths from single-dish emission measurements over the last two decades. This plot suggests that the systematic differences between the different methods are comparable to the uncertainties on individual measurements, although most DCF measurements are unfortunately given without formal uncertainties. Structure function measurements are typically amongst the higher magnetic field strengths, for a given density.

## 3.2. Comparison of DCF and Zeeman Measurements

The DCF method provides only an indirect measurement of magnetic field strength. The only direct method of measuring magnetic field strengths in molecular clouds is through Zeeman splitting of line emission from paramagnetic molecules (e.g., Crutcher, 2012). However, measuring the Zeeman effect in the environments of molecular clouds is extremely technically challenging, and unambiguous detections remain sparse (Crutcher, 2012), making indirect methods the only practical means of measuring magnetic fields in many ISM environments. However, in order to verify that the DCF method produces reasonable results, comparison must be made to Zeeman measurements where possible. Zeeman measurements of magnetic field strength are discussed in detail by Crutcher and Kemball (under review).

Crutcher et al. (2010) proposed an upper-limit relationship between gas density $n$ and total magnetic field strength $B$ in which $B = 10\mu$G at $n(H) < 300$ cm$^{-3}$, and $B \propto n^{0.65}$ at higher densities. This result was determined by Bayesian modeling of Zeeman measurements made in molecular clouds to that date. The estimated power-law index of 0.65 supports models in which the magnetic field is not strong enough to dominate over gravity in most environments considered. We show the Zeeman-derived Crutcher et al. (2010) $B - n$ relation (scaled to $n(H_2)$) on **Figure 2**. The DCF $B_{pos}$ measurements are broadly consistent with the Crutcher et al. (2010) relation, suggesting that DCF-derived field strength estimates are comparable to Zeeman measurements at these densities. While Crutcher et al. (2010) give their relation as an upper limit on magnetic field strength, some DCF-derived values significantly in excess of this relation are seen. These excesses could be caused by shortcomings in the DCF method, particularly failure to properly account for line-of-sight/sub-beam variations (cf. Hildebrand et al., 2009, although this method also gives values in excess of the Crutcher et al., 2010 relation), or by unaccounted-for uncertainties (as discussed by Pattle et al., 2017), or by genuinely highly magnetically-dominated systems (as claimed by, e.g., Pillai et al., 2015).

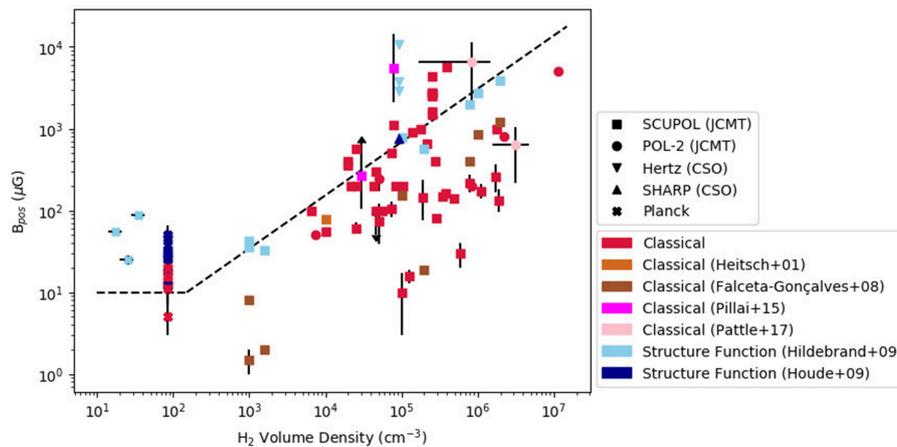

**FIGURE 2 |** A collation of DCF magnetic field strength measurements from single-dish emission polarimetry as a function of volume density. Note that some sources appear more than once on this plot. Dashed line marks the Crutcher et al. (2010) magnetic field/density relation, scaled to $n(H_2)$. All volume densities are converted to $n(H_2)$ if not given as such in the original work, assuming $n(H_2) = 0.5\,n(H)$, and mean molecular weight $\mu = 2.8$ (i.e., $n(H_2) = \frac{5}{6} n_{total}$), unless another value of $\mu$ is given in the original work. References: Davis et al. (2000), Henning et al. (2001), Matthews et al. (2002), Wolf et al. (2003), Vallée et al. (2003), Crutcher et al. (2004), Curran et al. (2004), Kirk et al. (2006), Curran and Chrysostomou (2007), Vallée and Fiege (2007), Hildebrand et al. (2009), Houde et al. (2009), Poidevin et al. (2013), Pillai et al. (2015), Planck Collaboration Int. XXXV (2016), Pattle et al. (2017), Kwon et al. (2018), Liu et al. (2018), Pattle et al. (2018), Soam et al. (2018), and Soler et al. (2018).

**Figure 2** shows that on average, DCF field strengths are comparable to those derived from Zeeman measurements. A more direct check on DCF results is comparison with Zeeman measurements in individual sources. However, such comparisons are complicated by the requirement that the species in which the Zeeman effect is observed traces the same material as the dust emission upon which the DCF analysis is performed (see discussion in Crutcher et al., 2004), and by the fact that DCF and Zeeman measurements trace orthogonal components of the magnetic field (cf. Heiles and Robishaw, 2009).

Ground-based submillimeter data sets typically trace volume densities $\geq 10^4\,\mathrm{cm}^{-3}$, and so CN and CCS are generally the only suitable tracers for direct comparison (cf. Crutcher et al., 1996). Comparisons of individual CN/CCS Zeeman measurements to SCUPOL-, Hertz-, SHARP-, and POL-2-derived DCF measurements generally find DCF field strengths to be the same order of magnitude as, but somewhat larger than, those determined from Zeeman measurements—see Kirk et al. (2006), Curran and Chrysostomou (2007), Hildebrand et al. (2009), Houde et al. (2009), Pillai et al. (2016), and Pattle et al. (2017).

Space-based and stratospheric instruments are less restricted in the volume densities which they can trace, and so allow direct comparison to the more easily detectable OH and H I Zeeman effects—for example, Soler et al. (2018) directly compare HI Zeeman measurements to Planck DCF measurements of the Eridanus superbubble, finding $B_{pos,DCF}/B_{los,HI} \sim 2.5 - 13$. In more distant and higher-mass regions, some comparisons can usefully be made between DCF measurements and Zeeman measurements from OH and $H_2O$ maser emission (e.g., Curran and Chrysostomou, 2007; Pattle et al., 2017), although care must be taken as maser emission traces only the extremely dense material surrounding the emitting source.

Poidevin et al. (2013) discuss comparison of CN and OH Zeeman and SCUPOL-derived DCF measurements in detail, finding that on average, $B_{pos}/B_{los} = 4.7 \pm 2.8$. They suggest several causes for this discrepancy: (1) line-of-sight field reversals to which Zeeman measurements are sensitive and DCF is not, (2) systematic differences in material traced, (3) the spatial averaging effects to which both methods are subject, (4) the possibility that DCF-inferred field strengths may be systematically overestimated due to integration effects (see discussion above), and (5) statistical differences between line-of-sight and plane-of-sky field strengths, noting that $B_{pos}$ will on average be larger than $B_{los}$, and a better tracer of total magnetic field strength (Heiles and Robishaw, 2009). Poidevin et al. (2013) argue that, in general, DCF-derived $B_{pos}$ provides an upper limit on the true magnetic strength, while Zeeman-measured $B_{los}$ provides a lower limit.

## 3.3. Intensity Gradient Technique

Koch et al. (2012a) proposed a method of measuring magnetic field strength based on the measured angle between magnetic field direction and gradient in emission intensity (see also Koch et al., 2012b, 2013). This method assumes that an emission intensity gradient is representative of the resultant direction of motion of material due to magnetic, pressure and gravitational forces. The "significance of the magnetic field"—the ratio of magnetic to gravitational and pressure forces—$\Sigma_B$, is given by

$$\Sigma_B \equiv \frac{F_B}{|F_G + F_P|} = \frac{\sin\psi}{\sin\alpha}, \quad (16)$$

where $\alpha$ is the angle between polarization direction and intensity gradient direction, $\psi$ is the angle between the direction of the local center of gravity and the intensity gradient direction, and $F_B$, $F_G$, and $F_P$ are the magnetic, gravitational and pressure forces, respectively. $\Sigma_B$ provides an estimate of whether the magnetic field is sufficiently strong to prevent gravitational collapse, with $\Sigma_B > 1$ indicating that the region under consideration is magnetically supported (Koch et al., 2012a).

Equation (16) can be rearranged to give magnetic field strength,

$$B = \sqrt{\frac{\sin\psi}{\sin\alpha}(F_G + F_P)4\pi R}, \quad (17)$$

where $R$ is the radius of curvature of the magnetic field. Note that this equation is given in cgs units.

This method requires estimation of a large-scale magnetic field curvature, and treats any deviation of polarization vector angles from the mean direction due to turbulent effects as a contaminant effect on the large-scale, ordered field. This method has the advantage of being able to provide a point-to-point estimate of magnetic field strength across a map, while DCF analysis provides an average value across a region. The method is also applicable to any measure of magnetic field direction, including, e.g., Faraday rotation. However, its applicability is limited to situations in which self-gravity is important, and is most applicable to the weak-field case in which magnetic fields are regulated by gravity (Koch et al., 2012a).

## 3.4. Velocity Gradient Technique

The velocity gradient technique (VGT; González-Casanova and Lazarian, 2017) indirectly estimates magnetic field strength and morphology in low-density, turbulent regions. VGT proposes that turbulence mixes magnetic field lines perpendicular to the local magnetic field direction, producing velocity gradients from which the magnetic field strength and morphology can be inferred. The VGT method works in simulations (González-Casanova and Lazarian, 2017), and predicts comparable magnetic field morphologies to those observed in dust polarization by Planck when applied to HI data (Yuen and Lazarian, 2017). This approach may usefully complement polarization measurements in the environments in which its assumptions can be expected to hold.

## 3.5. Histogram of Relative Orientations (HRO)

The Histogram of Relative Orientations (HRO; Soler et al., 2013) characterizes the dynamic importance of the magnetic field in molecular clouds through the distribution of angles between the projected magnetic field vectors on the plane of the sky and the column density gradient (indicative of the preferred direction of density structure) at each position. Simulations suggest that a weak magnetic field and/or non-self-gravitating density structure result in magnetic fields aligned along the preferred axis of the density structure, whereas a strong magnetic field and self-gravitating structure result in preferential alignment of high density structures orthogonal to the magnetic field direction (e.g., Soler et al., 2013; Wareing et al., 2016; Klassen et al., 2017). HROs provide a qualitative but powerful diagnostic of the relative importance of the magnetic field to a region. By restricting the analysis to progressively higher column densities, a threshold at which self-gravity becomes important can be identified (Chen et al., 2016).

The local orientation of cloud structure projected on the sky can be characterized by calculating the gradient field of the column density map $N_{\rm H}$. Since the gradient angle is normal to the local iso-$N_{\rm H}$ contour lines and the inferred magnetic field direction is perpendicular to the polarization orientation $\vec{E}$ the relative orientation angle between the local cloud orientation and the magnetic field is:

$$\phi = \arctan\left(|\nabla N_{\rm H} \times \vec{E}|, \nabla N_{\rm H} \cdot \vec{E}\right), \quad (18)$$

where $\phi$ is only unique within the range $0 \leq \phi \leq 90°$ (Soler et al., 2017). From this set of relative orientation angles a preference for parallel vs. perpendicular alignment can be calculated either from the HRO parameter,

$$\xi = \frac{N_{||} - N_{\perp}}{N_{||} + N_{\perp}}, \quad (19)$$

where $N_{||}$ and $N_{\perp}$ are the number of cloud sightlines where the local cloud orientation is parallel or perpendicular to the inferred magnetic field direction (Soler et al., 2013), or with the projected Rayleigh statistic,

$$Z_x = \frac{\sum_i^n \cos 2\phi_i}{\sqrt{n/2}}, \quad (20)$$

(Jow et al., 2018). For both relative orientation statistics $\xi$ or $Z_x > 0$ implies that the cloud structure is more often aligned parallel rather than perpendicular to the magnetic field, while $\xi, Z_x < 0$ indicates that the relative alignment is more often perpendicular than parallel.

## 3.6. Interpretation of Polarization Fraction

In the relatively low-density environments of molecular clouds and cores, it can reasonably be assumed that where a preferred polarization direction exists, it is perpendicular to the local magnetic field direction (Davis and Greenstein, 1951). Various theories of how this alignment comes about exist, although recently the radiative alignment torques (RAT) mechanism (Dolginov and Mitrofanov, 1976; Draine and Weingartner, 1996; Lazarian and Hoang, 2007) has become increasingly favored (Andersson et al., 2015). The analyses described above assume that grains are aligned with their long axes perpendicular to the local magnetic field, and that polarization observations accurately trace this alignment. It is of great importance to know the conditions under which this assumption holds.

Depolarization—a decrease in observed polarization fraction—is often seen toward high-column-density regions (e.g., Alves et al., 2014; Kwon et al., 2018). Depolarization could result from geometrical effects (vector cancelation of the magnetic field due to integration along the line of sight and/or within the telescope beam in the plane of the sky), or from grains becoming misaligned with the magnetic field at high optical depths. In the RAT paradigm, this would occur due to the increasing attenuation of short-wavelength radiation with increasing depth into the cloud preventing progressively larger grains from being aligned (e.g., Andersson et al., 2015).

Interferometric measurements indicate highly ordered magnetic field structures on small scales in protostellar systems (Hull et al., 2017). Thus, there are at least some circumstances in which grain alignment can persist to very high optical depths. However, interferometric measurements to date have focussed on systems with some internal source of radiation, the photons from which could aid the alignment of grains. Whether grain alignment persists to the centers of starless cores is not yet clear.

A useful measure of how well grains are aligned—on the assumption that the underlying magnetic field is linear—is polarization efficiency, $\epsilon_p$ (Whittet et al., 2008). For optically thin polarized emission, polarization efficiency is identical to polarization fraction (e.g., Jones et al., 2015). In extinction polarimetry, polarization efficiency is defined as polarization fraction normalized by optical depth in the relevant band (Andersson et al., 2015, and refs. therein).

Typically, $\epsilon_p \propto A_V^{-\alpha}$, with $0.5 \lesssim \alpha \lesssim 1$ (Alves et al., 2014, 2015; Jones et al., 2015), suggesting that grains become progressively less well-aligned at higher optical depths. There is some indication of a break in behavior at $A_V \sim 20$, beyond which the power-law index steepens significantly, suggesting very poorly-aligned or entirely unaligned grains (Jones et al., 2015), as shown in **Figure 3**. This would put an upper limit on the column densities at which dust polarimetry can trace magnetic fields. Fissel et al. (2016) explore this relation in detail using BLAST-Pol observations of Vela C, investigating the dependence of $P$ on both column density ($\propto A_V$) and dispersion in polarization angle. Fissel et al. (2016) also discuss the degeneracy between depolarization, sub-beam effects, and integration along long sight-lines in such analyses.

It should be noted that the standard selection of polarization vectors by their signal-to-noise ratio in polarization fraction, $p/\delta p$ may create bias in the recovered value of $\alpha$. The sensitivity of recent observations allows the implementation of Bayesian analyses of the $\epsilon_p$-$A_V$ relationship (Wang et al., 2018), which should better inform future discussions of the limits of applicability of submillimeter polarization observations.

## 4. OBSERVATIONS OF MOLECULAR CLOUDS

Molecular clouds represent the largest structures associated with star formation that may be gravitationally bound. In this section we discuss polarization observations on large scales (>1 pc) within clouds, and discuss what these data reveal about both the properties of cloud magnetic fields, and the role of the magnetic field in influencing cloud formation and evolution.

The importance of magnetic fields in molecular clouds is typically parameterized by two quantities. The first is the Alfvén Mach number $\mathcal{M}_A$, which is the ratio of the turbulent velocity to the Alfvén speed $v_A = |B|/\sqrt{4\pi\rho}$, where the ratio of turbulent to magnetic energy $E_K/E_B \approx \mathcal{M}_A^2$. The second is the mass-to-flux ratio $\Phi$, the ratio of the cloud mass to the maximum mass that be supported by the magnetic flux through the cloud against cloud self-gravity.

As discussed in section 2 observing magnetic fields within molecular clouds is challenging: the fraction of dust emission that is polarized is typically less than 10%, and in some cases can be less than 1% (Planck Collaboration Int. XIX, 2015). Observations with ground-based telescopes have been mostly limited to observing the bright, high-column density regions of molecular clouds (e.g., Matthews et al., 2009; Dotson et al., 2010), with the exception of the maps from the SPARO instrument, which produced the first large scale polarization maps covering entire giant molecular clouds (Li et al., 2006).

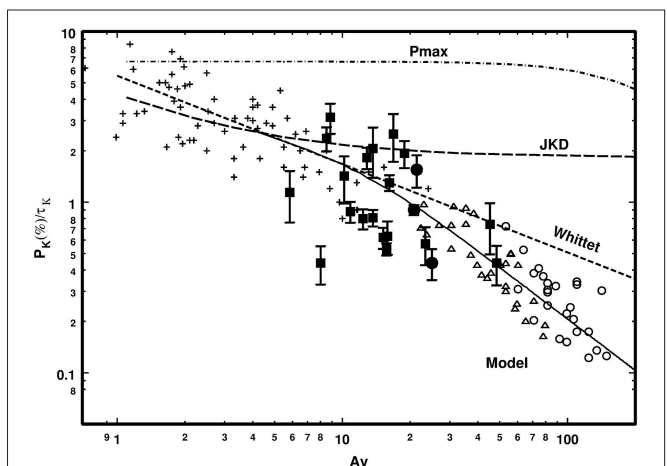

**FIGURE 3** | A plot of polarization efficiency ($\epsilon_p$), here defined as the K-band ratio of polarization fraction to optical depth ($p_K/\tau_K$), as a function of visual extinction $A_V$ in a starless core—Jones et al. (2015), Figure 5 © AAS. Reproduced with permission. Crosses and closed symbols show extinction polarimetric measurements; open symbols show submillimeter emission measurements. Note that polarization efficiency is defined in extinction as $\epsilon_p = p/\tau$ and in optically-thin emission as $\epsilon_p = p$; here, submillimeter emission points have been arbitrarily scaled to match extinction data. How effectively grains are aligned with the magnetic field is indicated by the gradient of the $\log \epsilon_p - \log A_V$ relation: the submillimeter data show a steeper gradient than the extinction data, suggesting that grains are less well-aligned at high optical depths.

A major recent advance is the release of all-sky polarization maps from the *Planck* satellite (Planck Collaboration Int. XIX, 2015). *Planck* had 4.8′ FWHM resolution at its 353 GHz (850 $\mu m$) frequency band, but due to sensitivity constraints *Planck* polarization maps typically require smoothing to at least 10′ FWHM resolution (∼0.4 pc resolution for the nearest molecular clouds at d ∼150 pc). In addition, the balloon-borne polarimeter BLASTPol (best FWHM resolution of 2.5′), has published extremely detailed polarization maps of the nearby giant molecular cloud Vela C at 250, 350, 500 $\mu m$ (Fissel et al., 2016; Soler et al., 2017). With the Planck and BLASTPol polarization maps it is now possible to apply the statistical analysis techniques discussed in section 3 to a large number of molecular cloud observations.

## 4.1. Where Does Polarized Dust Emission Best Trace Cloud Magnetic Fields?

Dust grains are thought to be aligned with respect to their local magnetic field due to radiative torques from relatively high energy photons (see discussion in section 3.6), and so the grain alignment is expected to be more efficient in the outer layers of molecular clouds. Since dust grains in the outer layers of clouds absorb more photons from the local interstellar radiation field (ISRF) they will also be warmer. These grains are therefore likely responsible for a larger fraction of both the total and polarized intensity, compared to colder, more shielded dust grains. Both of these factors imply that the magnetic field inferred from polarized dust emission is weighted toward the outer layers of molecular clouds.

Using BLASTPol 500 $\mu m$ observations of the Vela C giant molecular cloud Fissel et al. (2016) found that peaks in polarized intensity generally coincide with high column density regions, indicating that the polarization maps are tracing the cloud structure (see **Figure 4B**). However, polarization "holes" are observed toward several high column density regions. Fissel et al. (2016) modeled the decrease in fractional polarization with increasing column density, for the limiting case where all of the observed depolarization is caused by a decrease in grain alignment efficiency for more deeply embedded regions. They found that for moderate column density sightlines ($A_V \sim 10$) polarization measurements trace magnetic fields of all densities, while for high column density sightlines ($A_V \sim 40$) roughly half of the embedded dust contributes little to the polarized emission. This finding is in agreement with a study of polarization efficiency within a starless core by Alves et al. (2014), where it was found that grains remain largely aligned up to $A_V \sim 30$ mag.

Lower-resolution *Planck* maps of both molecular cloud envelopes and the diffuse ISM, also show regions of low polarization (Planck Collaboration Int. XIX, 2015). However, these observations can be explained entirely by turbulence and changes of the magnetic field geometry with respect to the line of sight, and do not appear to be caused by changes in grain alignment efficiency (Planck Collaboration Int. XX, 2015; Planck Collaboration et al., 2018).

As many techniques for analyzing magnetic fields using polarization data require comparison with simulations, it will be important to create realistic synthetic polarization maps from numerical simulations of star formation. Post-processing software such as POLARIS (Reissl et al., 2016), which can include calculations of grain alignment efficiency and variations in dust temperature, are now available. Seifried et al. (2019) applied POLARIS post-processing to the large scale SILCC-Zoom simulations, and found that the inferred magnetic field orientation from polarization maps with $\lambda \geq 160 \mu m$ typically agrees with the density averaged magnetic field angle to better than 10°.

## 4.2. Correlations Between Cloud Structure and Magnetic Field Orientation

Optical and near-infrared polarimetry observations show that high column density filamentary structures are often (e.g., B213 in Taurus, Goldsmith et al., 2008; Serpens South, Sugitani et al., 2011) but not always (e.g., L1495, Chapman et al., 2011) aligned perpendicular to the magnetic field. In contrast, lower density gas sub-filaments or "striations" seen in nearby clouds are often oriented parallel to the magnetic field direction (Goldsmith et al., 2008; Palmeirim et al., 2013).

Tassis et al. (2009) compared the orientation of cloud elongation to the inferred magnetic field orientation using archival CSO/Hertz polarization observations of structures ranging from nearby clumps to distant GMCs. They found a weak statistical preference for the cloud long axis to be oriented close to perpendicular to the magnetic field.

More recently, *Planck* and BLASTPol have produced large-scale polarization maps covering entire molecular clouds. These maps have been used to statistically analyze the relationship between cloud column density and magnetic field structure over many orders of magnitude in density and spatial scale. Planck Collaboration Int. XXXV (2016) studied 10 nearby clouds (d < 450 pc) with 10′ FWHM resolution (**Figure 5**) using the histogram of relative orientations (HRO) method discussed in section 3.5. They found that at low column densities ($N_H$) the cloud structure is more likely to align parallel than perpendicular to the magnetic field, while at $N_H$ greater than ∼$10^{22}$ cm$^{-2}$, the clouds structure is more likely to align perpendicular or have no preferred orientation to the magnetic field. This same trend is seen in synthetic polarization maps of numerical simulations only where the magnetic field is in equilibrium with or stronger than turbulence (Soler et al., 2013).

Using BLASTPol polarization maps of Vela C at 250, 350, and 500 $\mu m$ Soler et al. (2017) studied the relative orientation of the inferred magnetic field, which has resolution ∼0.6 pc, to the higher resolution (0.16 pc FWHM) column density maps derived from *Herschel* data. Similar to the results from Planck Collaboration Int. XXXV (2016), small-scale cloud structure of Vela C is preferentially aligned parallel to the cloud-scale magnetic field at low $N_H$, and perpendicular to the magnetic field at high $N_H$ (Soler et al., 2017; Jow et al., 2018). Furthermore, the slope of this transition from parallel to perpendicular is observed to be stronger in two cloud sub-regions dominated by dense, high-column density "ridge"-like structures, compared to the other two "nest"-like sub-regions where lower-$N_H$ filaments extend in many directions (**Figure 4**). Comparisons of the inferred magnetic field with orientation of structure in integrated molecular line intensity maps of Vela C show that low volume density molecular gas tracers (such as $^{12}$CO and $^{13}$CO) show structures aligned parallel to the magnetic field, while intermediate or high density gas shows a weak preference for alignment perpendicular to the large-scale magnetic field, with the transition occurring at $n_{H2} \sim 10^3$ cm$^{-3}$ (Fissel et al., 2018). These results show that in Vela C the cloud-scale magnetic field appears to have played an important role in the formation of small-scale and high density cloud sub-structure.

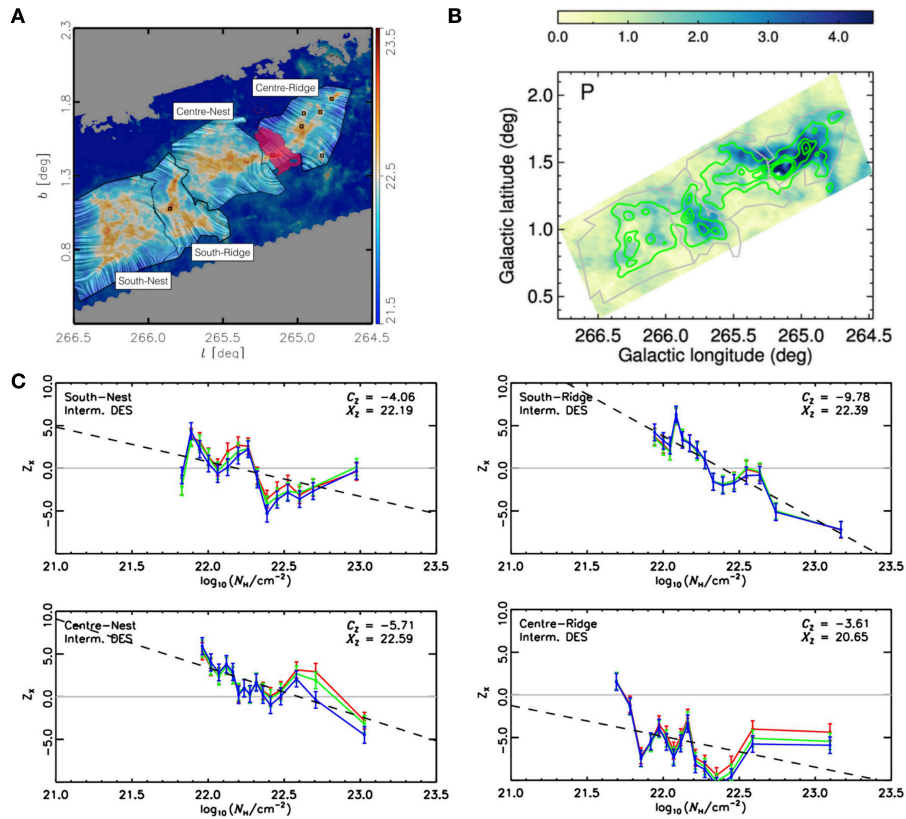

**FIGURE 4** | Polarization observations of the Vela C cloud from the BLASTPol telescope. **(A)** Magnetic field orientation (texture) inferred from the BLASTPol 500 $\mu m$ data overlaid on a *Herschel*-derived column density ($N_H$) map—Soler et al. (2017), reproduced with permission © ESO. Four sub-region are labeled, while the shaded pink region indicates where the dust is heated by a compact H$_{II}$ region. **(B)** Map of polarized intensity with contours of total intensity (green) overlaid (from Figure 3 of Fissel et al., 2016 © AAS. Reproduced with permission). **(C)** Relative orientation statistic $Z_x$ between the BLASTPol polarization orientation and the gradient of the *Herschel*-derived $N_H$ map, showing a transition from preferentially parallel orientation ($Z_x > 0$) in low $N_H$ regions, to perpendicular in high $N_H$ regions. Reproduced from **Figure 5** (Jow et al., 2018). By permission of Oxford University Press on behalf of the Royal Astronomical Society. This figure is not included under the CC-BY license of this publication[1].

Applying relative orientation analysis to synthetic polarization observations of numerical simulations, indicates that the slope and intercept of the relative orientation parameter, may encode information about the geometry of the flows that created the cloud (Soler and Hennebelle, 2017; Wu et al., 2017), or the magnetic field strength (Soler et al., 2013; Chen et al., 2016; Wu et al., 2017).

## 4.3. Magnetic Field Direction vs. Scale in Molecular Clouds

Molecular clouds are created from compressive flows in the more diffuse interstellar medium (ISM). One question of interest is whether magnetic fields preserve a "memory" of the local galactic magnetic field orientation. If the magnetic fields of molecular clouds are weak compared to turbulence then the field directions are expected to be decoupled from the field direction of the ISM.

Observations of the correspondence of molecular cloud fields and Galactic fields in the Milky Way are complicated by the long integration path of polarization observations through the Galactic disk. A study using CSO/Hertz polarization data by Stephens et al. (2013) found that while an average cloud magnetic field direction could be determined for most star forming regions (indicating relatively ordered fields), there was no clear correlation between the average cloud magnetic field direction and location on the Galactic plane, whereas the Galactic magnetic field is thought to be aligned parallel to the spiral arms (Heiles, 1996). However, many of the molecular clouds observed by Stephens et al. (2013) are high mass star forming regions, so the orientation of the magnetic field may have been modified by interactions with photo-ionized regions. Li and Henning (2011) compared the CO line polarization of six giant molecular clouds in the nearby galaxy M33 to the spiral arm orientation, and found a bi-modal relative orientation distribution consistent with alignment between the cloud magnetic field and the galactic field.

---
[1]Please visit: https://academic.oup.com/mnras/article/474/1/1018/4563620?searchresult=1. For permissions, please contact journals.permissions@oup.com

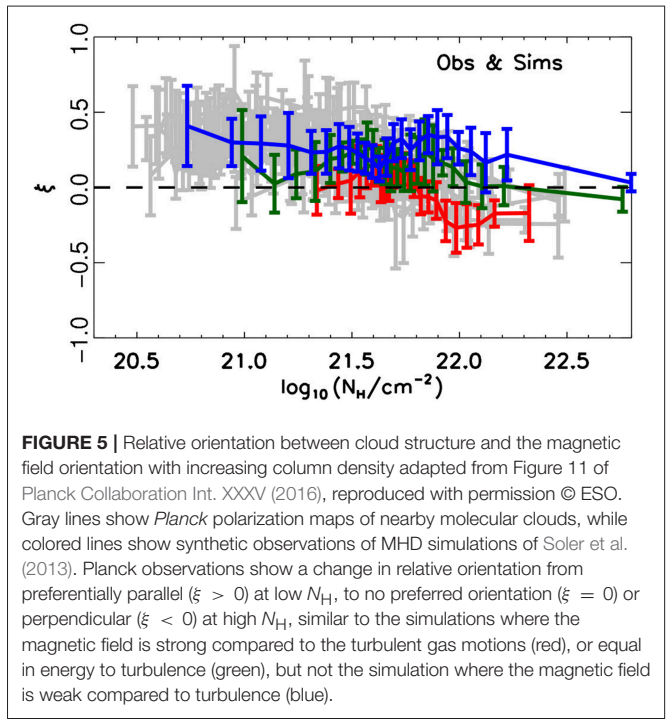

**FIGURE 5** | Relative orientation between cloud structure and the magnetic field orientation with increasing column density adapted from Figure 11 of Planck Collaboration Int. XXXV (2016), reproduced with permission © ESO. Gray lines show *Planck* polarization maps of nearby molecular clouds, while colored lines show synthetic observations of MHD simulations of Soler et al. (2013). Planck observations show a change in relative orientation from preferentially parallel ($\xi > 0$) at low $N_H$, to no preferred orientation ($\xi = 0$) or perpendicular ($\xi < 0$) at high $N_H$, similar to the simulations where the magnetic field is strong compared to the turbulent gas motions (red), or equal in energy to turbulence (green), but not the simulation where the magnetic field is weak compared to turbulence (blue).

In an earlier study by Li et al. (2009), the authors compared the orientation of the magnetic field in dense cloud sub-regions (1 pc or less) inferred from CSO/Hertz and JCMT/SCUPOL sub-mm polarization observations to the orientation of the magnetic field in the diffuse ISM surrounding the cloud from optical polarimetry. They found that 84% of all dense clumps have a difference in orientation of the clump vs. ISM field direction of less than 45°, and estimate that the probability of this occurring by chance is less than 0.01%. Attempts to reproduce this result in simulations by Li et al. (2015) indicate that the magnetic field in molecular clouds must be fairly strong; simulations with the magnetic field energy weaker than that of turbulence ($\mathcal{M}_A \gg 1$) cannot reproduce the correspondence between the observed core and large scale field direction.

## 4.4. Estimates of the Magnetic Field Strength Within Molecular Clouds

Estimates of magnetic field strength on cloud scales with the Davis-Chandrasekhar-Fermi (DCF) method discussed in section 3.1 are challenging because the available cloud scale sub-mm polarization maps from SPARO, BLASTPol, and *Planck* all typically have coarse resolution of several arcminutes. Any disordered field component on scales smaller than the telescope beam will not be observed, and this would lead to an overestimate of the POS magnetic field strength. Most estimates of large scale magnetic fields in molecular clouds with the DCF method use near-IR extinction polarimetry, since cloud envelopes typically have $A_V \ll 10$, such that background stars can still be observed (see for example Cashman and Clemens, 2014; Kusune et al., 2016).

SPARO observed four giant molecular clouds, with 4′ resolution, finding well ordered fields in two clouds, NGC6334 and G333.6-0.2, and two clouds where the magnetic field morphology appears to have been altered by feedback, the Carina Nebula and G331.5-0.1 (Li et al., 2006). Novak et al. (2009) used SPARO data and higher resolution CSO/Hertz follow-up observations to correct for the dispersion lost due to beam smoothing, and argue that the magnetic field strength must be at least as strong as turbulence in both NGC6334 and G333.6-0.2.

In Appendix D of Planck Collaboration Int. XXXV (2016), both the DCF method discussed in section 3.1.1 and the modified DCF modeling of the polarization structure function discussed in section 3.1.2 were applied to 10 nearby clouds observed with 10′ FWHM resolution *Planck* observations. Their estimated values of plane-of-sky magnetic field $B_{POS}$ range from 5 to 20 $\mu G$ for the DCF method alone, and 12 to 50 $\mu G$ using the DCF method combined with structure function analysis. Both methods of estimating magnetic field strength are consistent with mass-to-flux ratios $\Phi < 1$, which would imply that the magnetic field is strong enough to support the clouds against gravity. However, the authors of Planck Collaboration Int. XXXV (2016) caution that the measured dispersion in polarization angles is larger than the $\sigma_\theta \sim 25°$ threshold found in synthetic observations of numerical turbulence simulations by Ostriker et al. (2001), below which the DCF method can be used to estimate the magnetic field strength. They also note that the assumptions required for the structure function method of Hildebrand et al. (2009) and Houde et al. (2009) of a scale invariant random magnetic field component are not applicable to the Planck observations, and suggest that the values of magnetic field strength should be interpreted with caution.

## 4.5. Magnetic Fields in Photo-Ionized Regions

Giant molecular clouds often produce high mass stars, which then form photo-ionized regions that can alter both the structure of the parent cloud, and the morphology of the cloud magnetic field. Observations of magnetic fields in dense gas affected by feedback from high-mass stars remain scarce. Interpreting magnetic fields in such regions requires care, in order to distinguish between the effects of self-gravity and of external pressure on field geometry. For example the BLASTPol map of Vela C in **Figure 4B** shows a pinched field geometry toward the high density ridge associated with the cluster powering the RCW 36 HII region; however, it is unclear whether the field geometry is caused by a dragging of field lines by gravitational collapse or by the field geometry being shaped by the bipolar compact HII region (Soler et al., 2017).

The closest high-mass star-forming region, Orion, has been observed extensively with ground-based polarimeters. Polarization observations of Orion are discussed in detail in section 6.2.2.

Pattle et al. (2018) observed the photo-ionized columns (elephant trunks) known as the "Pillars of Creation" in M16 using POL-2 (**Figure 6**). They found that the field runs along the length of the Pillars, a morphology consistent with the field having been dragged or reorientated by the pillar formation process. However, the DCF estimated field strength is non-negligible (170–320 $\mu$G), sufficient to support the Pillars against radial collapse. This suggests that the process of pillar formation may have compressed an initially dynamically negligible field to be dynamically important, though the magnetic field is still not strong enough to prevent the destruction of the Pillars by the ionizing cluster.

Large scale *Planck* polarization observations of photo-ionized regions have also been used to learn more about the magnetic field properties in the host molecular cloud, and the characteristics of the compressed gas. Planck Collaboration Int. XXXIV (2016) used *Planck* polarization data for the Rosette Nebula in the Monoceros molecular cloud with Faraday rotation measurements to construct an analytic model of the magnetic field, where the magnetic field is inclined 45° to the line of sight with $|\vec{B}| = 6.5\text{–}9\,\mu$G in the molecular cloud.

## 4.6. Polarization Observations of Infrared Dark Clouds

Infrared Dark Clouds (IRDCs) are high column density, often filamentary, molecular clouds usually seen in silhouette against the bright IR emission of the Galactic plane, and may represent precursors of high mass star forming regions (Rathborne et al., 2006). Since IRDCs are typically > 1 kpc distant, they are typically studied with higher resolution ground-based polarimeters. Many IRDCs have been observed by JCMT/SCUPOL (Matthews et al., 2009), and CSO/Hertz (Dotson et al., 2010), however only a handful have been analyzed in any detail.

Pillai et al. (2015) analyzed JCMT/SCUPOL and CSO/Hertz observations of the IRDCs G11.11 − 0.12 and G0.253 + 0.016 (Matthews et al., 2009), finding that in G11.11−0.12 the magnetic field runs perpendicular to the main filament, while the field in a lower-density filament merging with the main filament is parallel to its length. In both IRDCs they infer that the energy in the magnetic field is at least as strong as energy of the turbulent motions of the gas ($\mathcal{M}_A \leq 1$), and comparable to that of self-gravity.

More recently Liu et al. (2018) observed the filamentary IRDC G035.39-00.33 with JCMT/POL-2, where the filament width was barely resolved with POL-2's $\sim 14''$ FWHM beam. Over most of the IRDC they found that the magnetic field is perpendicular to the main filament, with a DCF inferred magnetic field strength of $\sim 50\,\mu$G. However toward the massive collapsing starless clump candidate "c8," they infer a pinched magnetic field geometry implying that the field lines may be being dragged in by the infalling gas motions. Future observations with higher resolution JCMT/POL-2 450 $\mu$m polarimetry, IRAM/NIKA-2, or the upcoming LMT/TolTEC polarimeter (FWHM $\sim 5''$) may soon be able to resolve in detail the interaction between magnetic fields and gravity within nearby IRDCs.

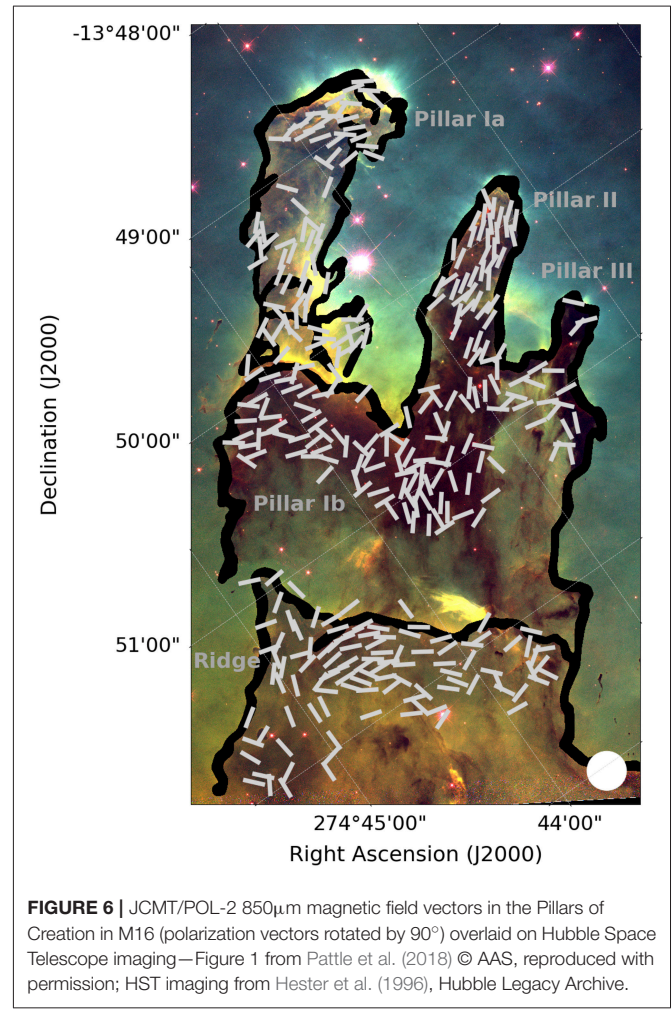

**FIGURE 6** | JCMT/POL-2 850$\mu$m magnetic field vectors in the Pillars of Creation in M16 (polarization vectors rotated by 90°) overlaid on Hubble Space Telescope imaging—Figure 1 from Pattle et al. (2018) © AAS, reproduced with permission; HST imaging from Hester et al. (1996), Hubble Legacy Archive.

## 5. MAGNETIC FIELDS IN Bok GLOBULES

Bok globules (Bok and Reilly, 1947), isolated clumps of molecular gas containing a few tens of solar masses within a diameter of a few tenths of a parsec (e.g., Launhardt et al., 2010), are a relatively simple environment in which the magnetic field geometry of starless and protostellar cores can be studied. As Bok globules are isolated objects (e.g., Alves et al., 2001), all emission associated with the globule is likely to come from the globule itself, although issues of grain misalignment at high densities remain (e.g., Jones et al., 2015). Bok globules may be starless or may harbor one or more protostars (e.g., Launhardt et al., 2010).

Submillimeter polarimetric observations to date have focussed on globules harboring protostars.

Most submillimeter polarimetric observations of Bok globules to date have been performed at 850 μm using SCUPOL, with which Vallée et al. (2000) observed CB 003, CB 034E, CB 054, CB 068, and CB 230, while Henning et al. (2001) observed CB 26, CB 54, and DC 253-1.6. CB 068 was marginally detected with the CSO/Hertz polarimeter (Dotson et al., 2010). Magnetic fields in Bok globules are generally found to be approximately linear in projection across the globule. Similarly, Ward-Thompson et al. (2009) observed the magnetic field across the Bok globule CB3 with JCMT/SCUPOL, finding the magnetic field to be linear in projection, and offset $\sim 40°$ to the core's minor axis—a likely projection effect (Basu, 2000; see also section 7).

Bok globules are an excellent environment for testing models of grain alignment, being isolated, fairly spherical, and generally having simple internal density structures and magnetic field geometries (e.g., Brauer et al., 2016). Depolarization at high column densities is typically observed in Bok globules (e.g., Vallée et al., 2000). However, at least some Bok globules show high polarization fractions at high densities, specifically CB 068, with $p \sim 10\%$ (Vallée et al., 2000). Vallée et al. (2003) argue that CB 068 (which hosts a young protostar) has an ordered field ($\sim 150 \mu G$), and low turbulence, making it a good environment for grain alignment to persist to high densities.

Wolf et al. (2003) estimated field strengths of $\sim 10^2 \mu G$ for the Bok globules B335, CB 230, and CB 244, all of which have embedded protostars. They find the magnetic field to be aligned with the major axis of B335 and CB 230, and compare these to the less evolved CB 26 and CB 54 (cf. Henning et al., 2001), in which the field is weakly aligned with the outflow axis. Wolf et al. (2003) propose that the magnetic field in such systems evolves from being aligned parallel with the outflow direction to being aligned parallel to the disc midplane.

## 6. MAGNETIC FIELDS WITHIN FILAMENTS

There is strong evidence for a bimodality in the orientation of magnetic fields with respect to filaments in molecular clouds (see section 4.2). Filaments are preferentially found to run either perpendicular or parallel to the local magnetic field direction in the surrounding, lower-density, medium. However, the behavior of magnetic fields within filaments is less well-characterized. In this section we summarize single-dish observations of magnetic fields within dense filaments, and in the immediate surroundings of filaments.

### 6.1. Magnetized Accretion Onto Filaments

André et al. (2014) proposed that filaments gain mass through magnetized accretion (see also Nakamura and Li, 2008; Palmeirim et al., 2013). In this paradigm, the sub-filaments, or striations, seen running perpendicular to self-gravitating filaments, and parallel to the magnetic field in the low density material surrounding these filaments, are accretion streams (Palmeirim et al., 2013). Star formation begins when the filament exceeds its maximum line mass for gravitational stability (Ostriker, 1964; see discussion below) and fragments.

Detections of magnetic fields running perpendicular to filaments on small scales have largely been made using optical or near-infrared extinction polarimetry (Sugitani et al., 2011; Palmeirim et al., 2013; Matthews et al., 2014; Panopoulou et al., 2016). Some submillimeter detections exist: Matthews et al. (2014) present BLAST-Pol observations marginally resolving the Lupus I filament, finding the magnetic field to run perpendicular to the filament direction, matching optical polarimetry results. Similarly, Cox et al. (2016) compare Planck 353 GHz observations of Musca to optical polarimetry and Herschel submillimeter imaging, finding the magnetic field in the low-density material to run perpendicular to the filament, and parallel to striations, as shown in **Figure 7**.

While Palmeirim et al. (2013) demonstrate large-scale red-shifted and blue-shifted CO emission preferentially located on opposite sides of the L1495 filament (using FCRAO data), the kinematics of such striations and sub-filaments—the theorized accretion streams—have not yet been observed in detail.

### 6.2. Magnetic Fields Within Nearby Filaments

The potential importance of magnetic fields within filaments was noted by Chandrasekhar and Fermi (1953). Magnetic fields may regulate the fragmentation and gravitational collapse of filaments (e.g., Fiege and Pudritz, 2000). However, the internal magnetic field geometry of filaments remains unclear. In order to conserve magnetic flux, field lines must either wrap around filaments (e.g., Nakamura et al., 1993; Fiege and Pudritz, 2000) or pass through them (e.g., Tomisaka, 2014; Burge et al., 2016).

Magnetic fields which wrap around filaments are referred to as "helical," loosely defined as a field with some form of toroidal and poloidal components. Such fields could be created through shear motion on an initially poloidal (axial) field (e.g., Fiege and Pudritz, 2000). Toroidal and poloidal fields play different roles in filament dynamics: poloidal fields provide support against collapse and fragmentation of filaments, while toroidal fields provide a confining tension (Fiege and Pudritz, 2000). Helical field geometries generally predict a decrease in polarization fraction toward the filament axis, an effect potentially degenerate with depolarization due to grain misalignment at high densities (e.g., Matthews et al., 2001b).

Magnetic fields which pass through a filament (generally referred to as "perpendicular") are expected to result in collapse of an initially cylindrical filament into a flattened, ribbon-like structure which may have an hourglass magnetic field across its cross-section (Tomisaka, 2014; Burge et al., 2016). In projection, the field lines will run along the length of a filament, with alternating minima and maxima in polarization fraction predicted across the filament's width (Tomisaka, 2015). Such a polarization structure has not yet been definitively observed, but provides a useful discriminant between the perpendicular and helical field models.

Observed filament radial density profiles may provide indirect evidence for the magnetic field geometry, and potentially a means of breaking projection effect and grain misalignment degeneracies. In unmagnetized filaments, density is predicted to fall as $r^{-4}$ in the filament wings (Ostriker, 1964). For purely poloidal fields, density is predicted to fall off faster than $r^{-4}$, while for generically helical fields, the predicted index is shallower, varying from $r^{-1.8}$ to $r^{-2}$ (Fiege and Pudritz, 2000). In models of perpendicular fields, the predicted index varies with model, but is shallower than the unmagnetized value (Tomisaka, 2014). However, all of these models are of non-accreting filaments, which is unlikely to be the case in practice. An understanding of the effect of accretion on observed filament density profiles would be necessary in order to use such profiles as a discriminant between magnetic field geometries.

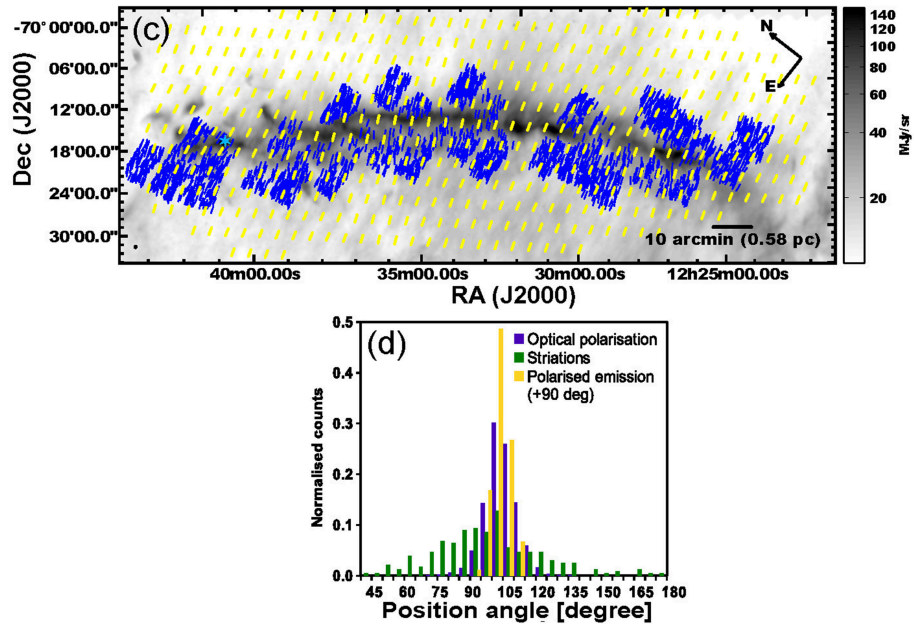

**FIGURE 7 |** The magnetic environment of the Musca filament; Figures 4c,d from Cox et al. (2016), reproduced with permission © ESO. **(Upper)** Image: Herschel SPIRE 250μm emission. Yellow vectors: Planck 353 GHz polarization vectors, rotated to trace the magnetic field direction. Blue vectors: starlight polarization vectors, tracing magnetic field direction. **(Lower)** Histograms of optical polarization, rotated emission polarization, and striation position angles, showing magnetic field direction and striation direction to be strongly peaked perpendicular to the direction of the filament.

Gravitationally unstable filaments are expected to fragment and collapse (Stodólkiewicz, 1963; Ostriker, 1964). Fiege and Pudritz (2000) presented a modification of the Ostriker (1964) critical line mass, taking into account magnetic support:

$$\left(\frac{M}{L}\right)_{crit,mag} = \left(\frac{M}{L}\right)_{crit}\left(1 - \frac{\mathcal{M}}{|\mathcal{W}|}\right)^{-1}, \qquad (21)$$

where $M$ is the mass of a filament of length $L$, $(M/L)_{crit} = 2c_s^2/G$, the Ostriker (1964) critical line mass ($c_s$ is sound speed, sometimes replaced with the full velocity dispersion), $\mathcal{M}$ is the magnetic energy per unit length, and $\mathcal{W}$ is the gravitational energy per unit length. $\mathcal{W} = -(M/L)^2 G$ for a generic uniform filament. Tomisaka (2014) also presents comparable magnetic critical line mass relations. In extremely massive filaments, the magnetic field may be distorted by flux-frozen gas motions caused by gravitational collapse of material along the filament (see discussion of OMC 1 below).

### 6.2.1. Planck Results

While Planck observations do not have sufficient resolution to observe fields within filaments in detail, Planck Collaboration et al. (2016) discuss their observations of the large-scale magnetic field morphology in three nearby filaments (Musca, L1506, B211), subtracting background emission by polynomial fitting. In these cases polarized emission from the filament can be separated from the "background" emission from the large-scale, low-density molecular cloud. The polarization angle in the filaments is found to be coherent, and offset from the background value by 12° (Musca), 54° (L1506), and 6° (B211; not significant), consistent with various models (e.g., Fiege and Pudritz, 2000; Tomisaka, 2014).

### 6.2.2. Filaments in Orion A and B
#### 6.2.2.1. Orion A OMC-1

The high-mass OMC-1 region within the nearby Orion A "integral filament" has been observed many times in polarized emission (Rao et al., 1998; Schleuning, 1998; Vallée and Bastien, 1999; Coppin et al., 2000; Houde et al., 2004; Hildebrand et al., 2009; Ward-Thompson et al., 2017). The mean magnetic field direction in OMC-1 differs significantly from that in the rest of the integral filament (Houde et al., 2004). While the average magnetic field direction in OMC-1 is approximately perpendicular to the direction of the filament (Ward-Thompson et al., 2017), the field shows significant ordered deviations from cylindrical symmetry, particularly in a large-scale "hourglass" feature (Rao et al., 1998; Schleuning, 1998) centered on the gravitational interaction between the Orion BN/KL and South clumps (shown in **Figure 8**). This field morphology is posited to result from motion of these two massive clumps along the filament under gravity (Schleuning, 1998; Pattle et al., 2017). The field is highly ordered and strong, with DCF-method-measured strengths varying from 0.76 mG (Houde et al., 2009) to 6.6 ± 4.7 mG (Pattle et al., 2017). The observed distortion of the field suggests that OMC-1 is not magnetically dominated, although energetics analysis suggests that the magnetic field may have been compressed to become dynamically significant (Pattle et al., 2017). The three-dimensional magnetic field geometry of the region is not clear; Schleuning (1998) propose a model in which the magnetic field passes directly through the filament at an angle, but large-scale helical geometries for the integral filament have also been proposed (e.g., Poidevin et al., 2011; Schleicher and Stutz, 2018).

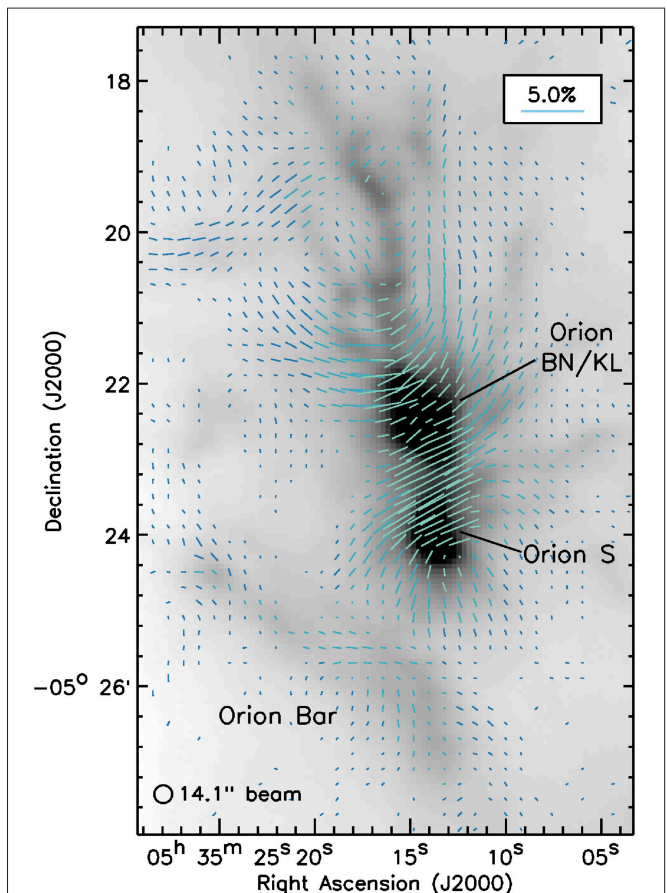

**FIGURE 8** | The magnetic field morphology in OMC 1; Figure 1 from Pattle et al. (2017) © AAS. Reproduced with permission. Figure shows JCMT/POL-2 850 μm polarization vectors, rotated to trace magnetic field direction, overlaid on a SCUBA-2 850 μm emission image. Note "hourglass" magnetic field morphology, centered on the interaction between the Orion BN/KL and S clumps.

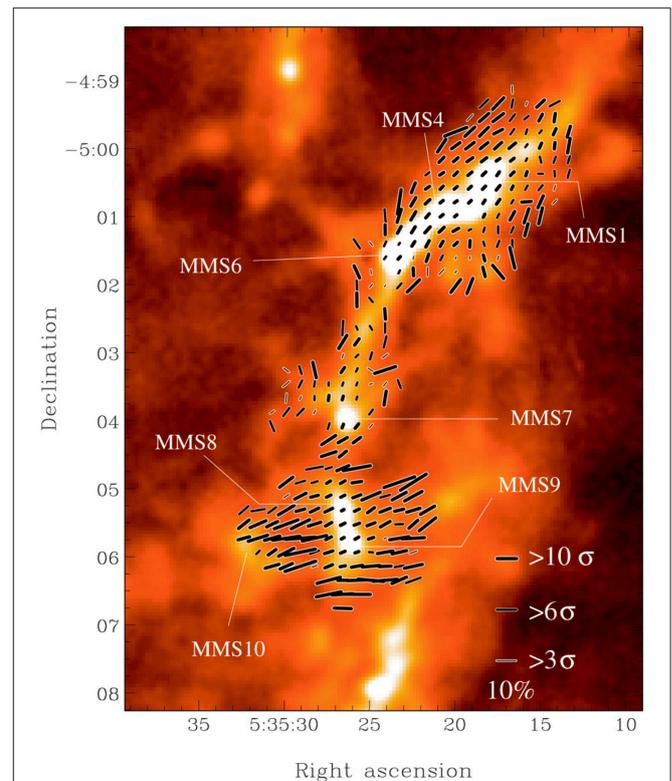

**FIGURE 9** | The 850 μm polarization geometry in OMC 3; Figure 2 from Matthews et al. (2001b) © AAS. Reproduced with permission. JCMT/SCUPOL 850 μm polarization vectors (note: not rotated), overlaid on SCUBA 850 μm intensity map. Note significant change in polarization direction along the length of the filament.

Monsch et al. (2018) observed a narrow, low-mass filament in the OMC-1 region in $NH_3$, and found that the magnetic field as observed with JCMT/POL-2 (Pattle et al., 2017; Ward-Thompson et al., 2017) runs parallel to filament. The filament has a very steep density profile, $r^{-5.1}$, inconsistent with predictions for toroidal fields but potentially consistent with an axial or perpendicular field. Both field and filament appear to extend radially from Orion BN/KL (the center of the OMC-1 region). It is thus a candidate for a "sub-filament," channeling material onto the central massive filament, in the André et al. (2014) model.

#### 6.2.2.2. Orion A OMC-3

OMC-3 is considerably less massive than OMC-1, and so the dynamics of the region are less gravity-dominated (e.g., Salji et al., 2015). Matthews et al. (2001b) observed several independent vectors over the width of the filament with JCMT/SCUPOL, finding that the magnetic field geometry is consistent with a toroidal field wrapping the filament along most of its length. These observations are shown in **Figure 9**. Houde et al. (2004) (CSO/Hertz, 350 μm) observed similar magnetic field geometries in OMC-3, but instead interpreted the magnetic field as perpendicular to the local filament direction. They also found that the average polarization direction remains approximately constant relative to a fixed position on the sky along the OMC-3 filament, rather than changing direction as the filament does, suggesting that the magnetic field is relatively unaffected by gravitational effects in the filament.

Matthews et al. (2001b) discuss the significant depolarization seen toward the filament axis at 850 µm, which could be due to decreasing grain alignment. However, they measure $\log p \propto -0.65 \log I$, which they note suggests grains are quite well-aligned at high densities, and that the observed depolarization is also consistent with predictions for a helical field. Matthews et al. (2001b) found no difference in behavior between regions of the filament with cores and those without, although the cores themselves are not well-resolved. Interferometric follow-up (Matthews et al., 2005) suggests that the field in the embedded cores is broadly aligned with the field in the larger filament.

Observations of OMC-3 provide a case study in the care which must be taken in the interpretation of polarized emission from objects that contain resolvable structure at many densities, in order to determine which structures the observed polarized emission is tracing. This is an issue on all size scales discussed in this work, but is particularly relevant to observations of filaments, in which the magnetic field properties may be expected to differ between the low-density envelope, the dense filament, and any embedded cores into which the dense filament has fragmented.

**Figure 10** (Li et al., in prep.) shows CSO/SHARP 350 µm (unpublished data) and JCMT/SCUPOL 850 µm observations of the OMC 3 region. Polarized intensity is shown in grayscale, with contours of total intensity overlaid. The JCMT/SCUPOL polarized intensity data show clear peaks associated with peaks in total intensity, whereas the shorter-wavelength CSO/SHARP data show no such correlation, with significant polarized intensity, but no clear peaks in emission. This suggests that the longer-wavelength JCMT/SCUPOL data are tracing the denser parts of the filament, while the CSO/SHARP data are tracing the filament envelope.

For optically thin emission, polarization observations not tracing the densest structure is likely to be due to grain misalignment at high densities (e.g., Jones et al., 2015), and a reasonable test of the densities traced by polarized emission is whether polarized intensity is correlated with total intensity. Where such a correlation exists, polarized emission can be expected to trace the full column of material. **Figure 10** can thus be interpreted as showing that in the dense parts of the filament, the larger, colder dust grains (which emit more of their light at longer wavelengths; e.g., Ossenkopf and Henning, 1994) are better aligned with the magnetic field than the smaller, warmer grains. In the RAT paradigm of grain alignment, the extinction of short-wavelength photons would prevent the alignment of small grains at high densities (e.g., Andersson et al., 2015). Thus the 350 µm data traces the magnetic field in the envelope surrounding the filament, while the 850 µm data traces the dense material of the filament itself. This may not be the case in all filaments, as the densities traced at a given wavelength will depend on grain properties, temperature and interstellar radiation field (ISRF). An additional caveat is that the difference in the chop throws of the SHARC-II and SCUBA cameras on detectable size scales has not been fully explored (see Fissel et al. (2016) for a discussion of the effect of background subtraction on polarization observations). Although this source provides an illustrative example only, consideration of such correlations is likely to be of general use.

#### 6.2.2.3. Orion A OMC-2
Houde et al. (2004), using CSO/Hertz, found that the magnetic field in the north of OMC-2 agrees with that in OMC-3, but changes abruptly in the south of the region. Houde et al. (2004) tentatively associate this change with outflow activity in the vicinity of the source OMC-2 FIR 3. However, Poidevin et al. (2010), observing with JCMT/SCUPOL, did not find a correlation between outflow direction and magnetic field direction. Poidevin et al. (2010) found that OMC-2 is more weakly polarized than OMC-3, with a steeper decrease of $p$ with $I$, a less well-ordered magnetic field, and higher levels of turbulence. Poidevin et al. (2010) argue that while magnetism dominates over turbulence in OMC-3, this is not clearly the case in OMC-2.

#### 6.2.2.4. Orion A OMC-4
Houde et al. (2004) observed a small number of vectors in the OMC-4 region, finding a field orientation not clearly related either to that of the larger integral filament or to the geometry of the OMC-4 region itself.

OMC 1–4 are all contiguous parts of the integral filament (e.g., Bally, 2008). The magnetic field apparently having a different geometry and dynamic importance in different parts of the filament—with OMC-1 gravitationally-dominated, OMC-2 turbulence-dominated, and OMC-3 magnetically-dominated—suggests that the behavior of magnetic fields within filaments, and the evolution of the filaments themselves, depends strongly on local as well as large-scale environment.

#### 6.2.2.5. Orion B
Matthews et al. (2002) observed the NGC 2071 and LBS 23N cores (discussed in section 7) and the NGC 2024 filament in Orion B at 850 µm using JCMT/SCUPOL. NGC 2024 shows an ordered polarization geometry, consistent with a toroidal field threading the filament. They alternatively model the field toward NGC 2024 as resulting from the sweeping up of dense, magnetized gas by a foreground H II region (see also section 4.5), with the filament itself unmagnetized, but conclude through qualitative comparison with models that a helical field geometry within the filament is more consistent with observations. BIMA follow-up of NGC2024 shows that the small-scale field in the embedded cores generally matches that of the filament (Lai et al., 2002), supporting this interpretation.

### 6.3. Future Directions
In order to further our understanding of the magnetic field geometry within filaments, it is necessary to break the degeneracy between depolarization due to geometrical effects and that due to grain (mis)alignment at high densities. This requires observations with sufficient sensitivity and resolution to observe good radial profiles of polarization fraction and polarized intensity across filaments, as well as detailed predictions of polarization fraction as a function of radius for the various proposed field geometries. Multi-wavelength observations can be used to investigate dust properties and the depth into filaments traced by magnetic fields, thus allowing some quantification of the reliability of polarization fraction as a tracer of field geometry.

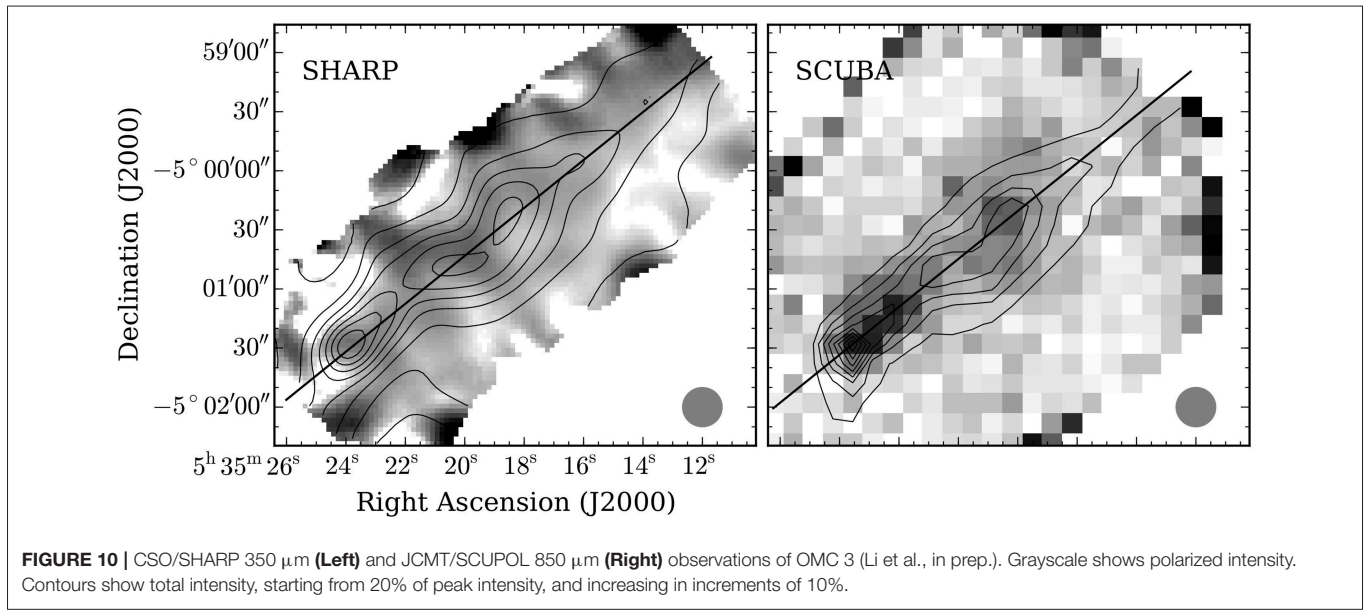

**FIGURE 10** | CSO/SHARP 350 µm **(Left)** and JCMT/SCUPOL 850 µm **(Right)** observations of OMC 3 (Li et al., in prep.). Grayscale shows polarized intensity. Contours show total intensity, starting from 20% of peak intensity, and increasing in increments of 10%.

## 7. POLARIZATION OBSERVATIONS OF STARLESS AND PROTOSTELLAR CORES

We here define starless cores to be small-scale overdensities within larger molecular clouds which, if gravitationally bound, will form an individual star or system of stars (Benson and Myers, 1989). Prestellar cores are the gravitationally bound subset of starless cores (Ward-Thompson et al., 1994). Protostellar cores are defined as envelope-dominated sources containing one to a few hydrostatic objects (i.e., Class 0 and I sources; Lada, 1987; Andre et al., 1993).

Detections of starless and protostellar cores in polarized light have until recently been piecemeal, and strongly limited by surface brightness. It is now becoming possible to systematically survey nearby star-forming regions to map magnetic fields in starless and protostellar cores. Total-power instruments remain the best tools for detecting starless cores, while protostellar cores are now more commonly observed with interferometers.

### 7.1. Starless Cores

The number of instruments with both the sensitivity and resolution required to detect polarized emission from starless cores remains very limited. Starless cores are extended objects with simple internal geometries, typically well-modeled by Bonnor-Ebert (Ebert, 1955; Bonnor, 1956) or Plummer-like (Plummer, 1911) distributions (e.g., Alves et al., 2001; Whitworth and Ward-Thompson, 2001), making observations with a total power component essential, as interferometers typically resolve out starless cores entirely.

Due to their small size and low surface brightness, imaging of individual starless and prestellar cores is largely restricted to the most local star-forming regions. The first detection of polarized submillimeter emission from three dense starless cores was reported by Ward-Thompson et al. (2000), who observed L1544 (140 pc; Elias, 1978), L183 (180 pc; Ward-Thompson et al., 2000) and L43 (150 pc; Ward-Thompson et al., 2000) at 850 µm using JCMT/SCUPOL. Crutcher et al. (2004) used the DCF method to estimate magnetic field strengths for the same sources, finding $B_{pos} = 140\,\mu$G, $80\,\mu$G and $160\,\mu$G in L1544, L183, and L43, respectively, and that the three cores were, after correction for geometrical bias, approximately magnetically critical.

Kirk et al. (2006) observed two less-dense starless cores, L1498 and L1517B (both 140 pc), with JCMT/SCUPOL, estimating plane-of-sky field strengths of $10 \pm 7\,\mu$G and $30 \pm 10\,\mu$G respectively, again using the DCF method. The former value is comparable to a line-of-sight Zeeman measurement of the same region ($48 \pm 31\,\mu$G; Levin et al., 2001). In these cores, thermal support was found to dominate over non-thermal and magnetic support, with the cores being magnetically supercritical (unable to be supported by their internal magnetic fields alone).

Magnetic fields detected in isolated starless cores are typically relatively smooth and well-ordered, with detectable polarization across the cores. An example of such a field, in the starless core L183, is shown in **Figure 11**. Ward-Thompson et al. (2000) found that magnetic fields over the central core regions are typically aligned at $\sim 30°$ to the projected minor axis of the cores, a result ascribed to projection effects by Basu (2000). Notably, despite their ordered field morphologies, and despite being candidates for gravitational instability (e.g., Kirk et al., 2006), none of these cores show the classical "hourglass" magnetic field characteristic of ambipolar-diffusion-driven collapse. The precise role of the magnetic field in the evolution of these isolated cores is not clear. However, the magnetic field does not appear to be dynamically negligible, particularly in the denser set of cores.

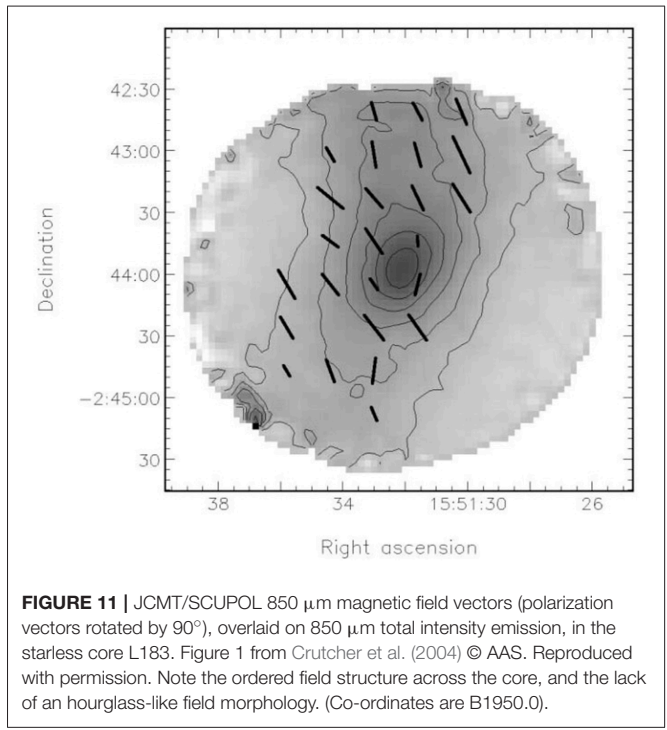

**FIGURE 11** | JCMT/SCUPOL 850 µm magnetic field vectors (polarization vectors rotated by 90°), overlaid on 850 µm total intensity emission, in the starless core L183. Figure 1 from Crutcher et al. (2004) © AAS. Reproduced with permission. Note the ordered field structure across the core, and the lack of an hourglass-like field morphology. (Co-ordinates are B1950.0.)

The full set of observations made with JCMT/SCUPOL are cataloged by Matthews et al. (2009). This includes the five starless cores described above and L1287, observed by Curran and Chrysostomou (2007), listed as starless by Matthews et al. (2009), but with an associated energetic outflow (Curran and Chrysostomou, 2007). The JCMT/SCUPOL archive also contains observations of several nearby star-forming regions within which individual cores can be resolved, particularly the L1688 region in Ophiuchus: Oph A (Tamura, 1999), Oph B2 (Matthews et al., 2001a), and Oph C (Matthews et al., 2009).

Alves et al. (2014) observed the starless core Pipe-109 with APEX/PolKa, finding a highly-ordered magnetic field with significant depolarization at high column densities (note also Alves et al., 2015).

Observations made using CSO/Hertz are cataloged by Dotson et al. (2010). This catalog contains no isolated starless cores, but includes the Oph A clump, containing a number of embedded starless cores.

A number of regions containing starless cores have been observed by the JCMT/POL-2 polarimeter, with significantly better sensitivity than was possible with its predecessor, SCUPOL. The Oph A and B clumps have been observed by Kwon et al. (2018) and Soam et al. (2018), respectively, as part of the BISTRO survey (Ward-Thompson et al., 2017). The $\sim$ 1800 AU linear resolution of these observations permits insight into the magnetic field morphology in the many starless and protostellar cores within the clumps (cf. Motte et al., 1998; Pattle et al., 2015). Kwon et al. (2018) measure field strengths varying from $0.2 - 5$ mG across Oph A, suggesting that the magnetic field is unlikely to be dynamically negligible anywhere in the region, but may vary significantly within it. Soam et al. (2018), observing the Oph B1 and B2 clumps, infer a typical magnetic field strength in Oph B2 of $630 \pm 410$ µG, again suggesting that the magnetic fields in the cores in the region will not be negligible.

Discussion of these observations of well-resolved clumps has largely focussed on the properties of the clumps themselves, rather than individual embedded cores, in part due to limitations in sensitivity, but also due to uncertainties as to whether polarized emission from clumps and filaments traces the dense, star-forming gas or the larger- (clump/filament-)scale material (see discussion in section 6). Oph A has recently been observed in the far-infrared with the SOFIA/HAWC+ polarimeter (Harper et al., 2018; Santos et al., in prep.). These observations are shown alongside the JCMT/POL-2 polarization map of the region in **Figure 12**. Forthcoming polarization spectra across the $1-1,000$ µm wavelength regime will provide additional insight into grain physics in regions such as Oph A, as discussed in section 2.

## 7.2. The Search for High-Mass Prestellar Cores

The debate over the existence of high-mass prestellar cores (with masses several times their Jeans mass, collapsing to form a massive star; e.g., Tan et al., 2014) continues. If such objects exist, they are likely to require significant internal magnetic support (e.g., Pillai et al., 2015). Pillai et al. (2015) analyse JCMT/SCUPOL observations of G11.11-0.12, positing that the source is a magnetically supported high-mass starless core. Due to the distance of most high-mass star-forming regions, most detections of high-mass star-forming "cores" are interferometric, for example, polarimetric observations of high-mass cores in DR21 (Ching et al., 2017), and in W51 (Tang et al., 2013) taken using the SMA. Single-dish data provide context for these observations, by mapping the larger-scale magnetic field in the surrounding material (e.g., Dotson et al., 2000; Chrysostomou et al., 2002; Vallée and Fiege, 2006).

## 7.3. Protostellar Cores

Protostellar cores differ from starless cores in that they have an internal heating source, and potentially an internal source of ionizing photons. These cores are thus generally warmer and brighter than their starless counterparts, and may be expected to contain dust grains better aligned with their internal magnetic fields (Jones et al., 2016).

Thanks to the presence of embedded sources and complex internal gas and dust structures (discs, accretion flows, etc.), protostellar cores make excellent targets for interferometric polarimetry. Much of the focus of polarimetric studies of protostellar cores has shifted to interferometric measurement— such as imaging of hourglass magnetic fields in NGC 1333A using the SMA (Girart et al., 2006), large-scale surveys such as TADPOL with CARMA (Hull et al., 2014), and high-resolution imaging of complex magnetic field geometries around protostars with ALMA (Hull et al., 2017). We summarize single-dish observations of protostellar sources to date here.

Models of magnetic fields in protostellar environments generally predict a symmetric field about the outflow axis of the system, with net polarization aligned either along the direction of the outflow, or with the plane of the disc (e.g., Greaves et al., 1997). Early observations of the magnetic fields in the envelopes of young protostars were largely made using the Aberdeen/QMW polarimeter on the UKT14 camera on the JCMT (Minchin et al., 1995; Tamura et al., 1995; Holland et al., 1996; Greaves et al., 1997). Tamura et al. (1995) and Minchin et al. (1995) found magnetic fields perpendicular to the major axes of protostellar envelopes and aligned with large-scale molecular outflows, whereas Holland et al. (1996) found that in the prototypical Class 0 source VLA 1623, the magnetic field is perpendicular to the outflow, and so suggested that the outflow cannot be magnetically collimated on large scales. Greaves et al. (1997) found that for Class 0 sources with outflows aligned in or near the plane of the sky, the magnetic field is typically perpendicular to the outflow, whereas for outflows aligned close to the line of sight, the magnetic field is parallel to the outflow direction, a bimodality in behavior common to a number of models of the magnetic field structure of protostellar envelopes. Greaves et al. (1997) also found polarization fraction to be anticorrelated with outflow opening angle and with ratio of bolometric luminosity to 1.3 mm luminosity (both proxies for age), leading them to suggest that magnetic fields are more ordered in younger protostellar sources.

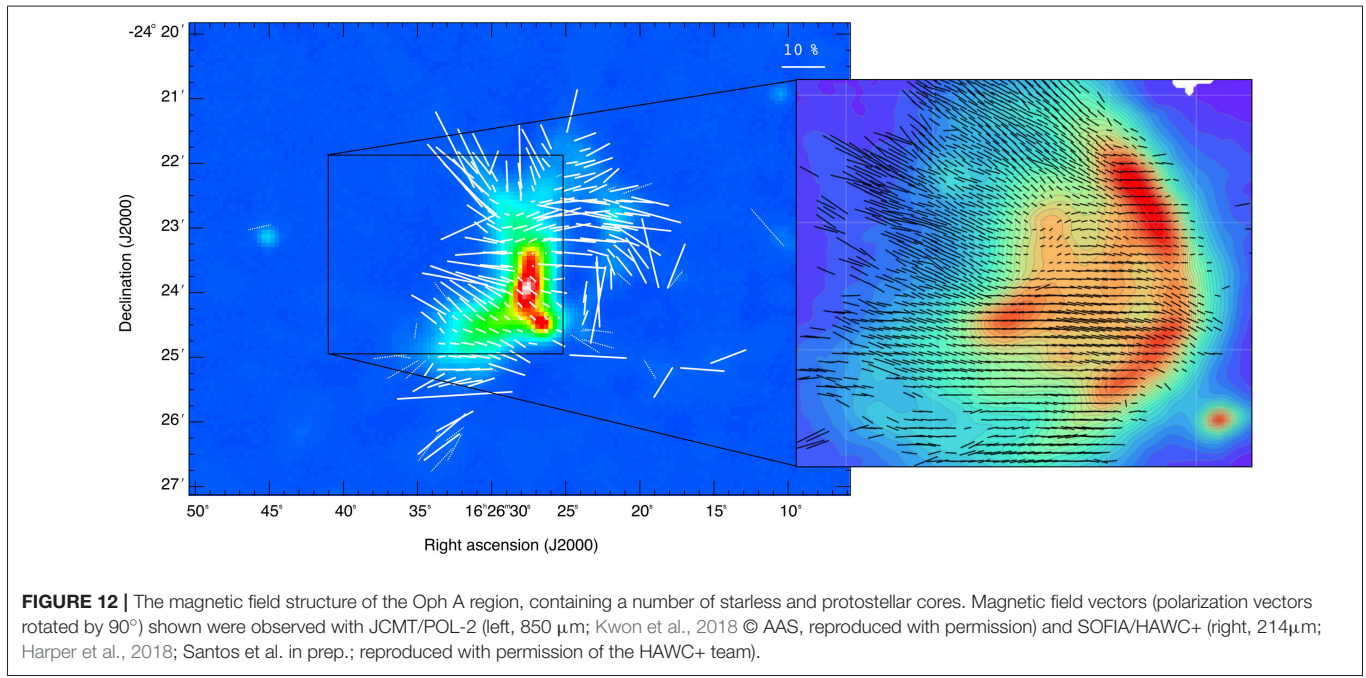

**FIGURE 12** | The magnetic field structure of the Oph A region, containing a number of starless and protostellar cores. Magnetic field vectors (polarization vectors rotated by 90°) shown were observed with JCMT/POL-2 (left, 850 µm; Kwon et al., 2018 © AAS, reproduced with permission) and SOFIA/HAWC+ (right, 214µm; Harper et al., 2018; Santos et al. in prep.; reproduced with permission of the HAWC+ team).

JCMT/SCUPOL observed a larger set of protostars and clumps containing embedded protostars, most of which were first published by Matthews et al. (2009). A number of protostellar sources have been observed using CSO/SHARP: NGC 1333 IRAS 4 (Attard et al., 2009), the Class 0 protostars B335, L1527, and IC348-SMM2 (Davidson et al., 2011), DG Tau (T Tauri) (Krejny et al., 2011), and the Class 0 protostar L1157-mm (Stephens et al., 2013). Of these Attard et al. (2009) and Stephens et al. (2013) found the magnetic field to be broadly aligned with outflow direction, while Davidson et al. (2011) found magnetic field vectors consistent with a pinched magnetic field geometry in the protostellar discs of B335 and IC348-MM2 (indicative of magnetized disc formation), while in L1527 they propose that the outflow is sufficiently energetic to have disordered the magnetic field. Chapman et al. (2013) stacked observations of seven protostellar cores observed with CSO/SHARP, and found a positive correlation between the magnetic field direction and the symmetry axis of the (stacked) core, as well as between the field and outflow directions. Chapman et al. (2013) also present some evidence for a pinch in the stacked magnetic field, consistent with magnetically-dominated evolution.

Single-dish polarimetric observations of clumps containing protostellar cores include observations of the Oph A region, discussed above. Other observations include the Orion B NGC 2071 and LBS 23N clumps (Matthews et al., 2002). NGC 2071 is a massive core forming multiple protostars, showing a uniform magnetic field geometry oriented perpendicular to the most powerful outflow in the region, with a DCF-inferred field strength of 56 µG. LBS 23N, however, shows a more disordered field geometry and significant depolarization toward its various starless and protostellar cores. Matthews and Wilson (2002b) observed the lower-mass Barnard 1 clump in Perseus, which again contains both protostellar and starless cores. An ordered polarization pattern is seen across the clump, with significant depolarization toward the dense cores. These observations of clumps containing starless and protostellar cores do not show significant differences in polarization fraction or $\log p - \log I$ index between starless and protostellar cores, on scales observed by the JCMT.

### 7.4. Comparison With Simulations
Few direct comparisons have been made between observations and simulations of magnetic fields in starless cores, in part due to the paucity of observations. Most comparison between models and observations has been to numerical and (semi-)analytic models of ambipolar-diffusion-driven core collapse (e.g., Mouschovias, 1976a,b; Ciolek and Mouschovias, 1994; Basu, 2000; Ciolek and Basu, 2000).

MHD modeling of star-forming cores began with Machida et al. (2004) and subsequent papers, which focussed on cores harboring protostars. Subsequent work includes, e.g., Dib et al. (2007) (clumps/cores), Price and Bate (2007) (protostellar environments), Boss and Keiser (2013) (protostellar environments and discs). These simulations have generally focussed on the time evolution of core collapse rather than on producing synthetic observations. Mocz et al. (2017) produce volume-weighted magnetic field maps for collapsing cores in a turbulent medium which, while presented at resolutions observable by interferometric instruments, could be smoothed to be comparable to maps of cores produced by single-dish instrumentation.

Radiative transfer modeling allows forward modeling of the magnetic fields observed in star-forming cores, using tools such as DustPol (Padovani et al., 2012) and POLARIS (Reissl et al., 2016) to produce magnetic field maps for model cores, to be compared to observations. Alternatively, modeling of observed field geometries can be treated as a highly degenerate inversion problem. POLCAT (Franzmann and Fiege, 2017) produces simulated polarization maps based on models of three-dimensional cores threaded by magnetic fields, in order to eliminate core and field geometries not consistent with the magnetic field observed in projection.

## 7.5. Depolarization

As discussed in section 3.6, the alignment of grains with magnetic fields is, in the absence of a source of energetic photons, expected to get progressively worse at increasing $A_V$. In the extreme case of a deeply embedded starless/prestellar core, it is not clear whether dust grains are coupled to the magnetic field (Jones et al., 2015).

Polarization holes—decreased polarization fraction at increased total intensity (typically a tracer of density in cold cores)—are seen in observations of starless cores (Ward-Thompson et al., 2000; Alves et al., 2014), of cores with very young hydrostatic sources (Kwon et al., 2018, observing the core SM1, which may contain an extremely young Class 0 source; Friesen et al., 2014), and in single-dish observations of sources with embedded protostars (Wolf et al., 2003; Jones et al., 2016).

Jones et al. (2016) present CSO/SHARP total power data alongside CARMA interferometric imaging of an IRDC with an embedded massive protostar. The total power data, tracing larger size scales in the IRDC, show a steeper negative slope in the $\log p$-$\log I$ relation than the interferometric data tracing the material around the embedded source. This is interpreted as evidence that although dust grains in the IRDC are in general not aligned with the magnetic field at high densities, photon flux from the young protostar is driving grain alignment in its vicinity.

The extent to which polarization holes results from grain misalignment, or from complex field morphologies on scales smaller than the beam, is not clear. The facts that starless cores and protostellar cores show similar behaviors on large scales (e.g., Matthews and Wilson, 2002b), that $\log p$-$\log I$ indices vary within and between clouds (Matthews and Wilson, 2002a; Matthews et al., 2002), and that ordered fields are consistently seen in starless cores despite depolarization (e.g., Kirk et al., 2006), suggest that better modeling of 3D magnetic field geometries is required in order to determine the depth into starless cores to which single-dish polarization observations can trace. Such modeling is becoming possible through application of tools such as POLARIS, as described above (e.g., Valdivia et al., 2017).

## 7.6. Future Directions and the Potential of Large Surveys

There are many open questions in the field of polarimetry of starless and protostellar cores, not least as to the circumstances in which it can be said with confidence that the magnetic field in the densest, gravitationally unstable, parts of cores has been observed. The new generation of polarimetric surveys will allow us to begin addressing these questions in a systematic manner, by allowing comparison of meaningful numbers of starless, protostellar, embedded and isolated sources. Wide-area surveys also allow the possibility of stacking polarization fraction and polarized intensity images to improve signal-to-noise and so to determine something of the magnetic field properties in low-surface-brightness starless cores.

## 8. SUMMARY

In this chapter we have discussed submillimeter and far-infrared single dish continuum emission polarimetric observations of magnetic fields in star-forming regions. We discussed strategies for measuring polarized dust emission, and reviewed past, present and upcoming polarimeters.

We then summarized the most widely-used methods for estimating the strength and dynamic importance of magnetic fields in molecular clouds, as well as the means by which the depth into clouds to which polarization observations trace can be assessed. We compared the various implementations of the (Davis-)Chandrasekhar-Fermi (DCF) method of estimating magnetic field strength. Compilation of DCF measurements shows that the DCF method typically produces magnetic field strength values comparable to those measured directly from Zeeman splitting of spectral lines, for a given density. There is considerable variation in DCF results, with differences in results between different DCF implementations typically comparable to or greater than measurement uncertainty. We note the importance of testing DCF and other experimental methods against synthetic observations in order to determine their validity and accuracy.

Our ability to study magnetic fields on molecular cloud scales has been revolutionized by all-sky observations from the *Planck* satellite, as well as cloud-scale maps from BLASTPol and SPARO. These observations indicate that the energetic importance of magnetic fields on $>1$ pc scales are typically equal to or larger than that of turbulent gas motions, and that magnetic fields appear to play an important role in the formation of dense cloud substructures. Observations of more clouds at higher resolution are needed to better constrain the role played by magnetic fields in all stages of molecular cloud formation and evolution.

Observations of Bok globules typically show ordered, linear, magnetic fields with field strengths $\sim 10^2\,\mu$G. Most Bok globules show significant depolarization at high densities. In Bok globules harboring outflow-driving sources, the magnetic field may in some cases be aligned with the outflow direction. Being by definition isolated objects, generally with simple geometries, Bok globules are a useful environment in which to study magnetic fields.

As magnetic fields tend to be perpendicular to self-gravitating filaments in the low-density environment surrounding the filaments, some models predict that material is accreted onto such filaments along these magnetic field lines—a theory with some observational support in nearby filaments such as Taurus and Musca. However, the three-dimensional magnetic field geometry within such star-forming filaments is not well-characterized. Magnetic fields are expected to either wrap helically around filaments or to pass directly through them. In order to distinguish between these alternatives, and to break degeneracies between three-dimensional geometry and grain misalignment, well-resolved observations across filaments are required, ideally at more than one wavelength. Care needs to be taken to ensure that polarization observations trace the dense material of filaments, rather than the low-density envelope. The role of magnetic fields within filaments is likely to vary significantly with environment: for example, the well-studied integral filament in Orion A shows gravitationally-dominated, turbulence-dominated and magnetically-dominated behavior at various points along its length.

Observations of magnetic fields within starless cores remain strongly limited by surface brightness. Where detectable, magnetic fields in isolated starless cores are typically relatively smooth and well-ordered, with polarization detected across the cores, although depolarization toward high densities is seen. While observations of magnetic fields in starless cores do not clearly show the "hourglass" morphology predicted for magnetically-dominated, ambipolar-diffusion-driven star formation, the ordered polarization patterns seen suggest that magnetic fields are of some dynamic importance in these objects. The depth into starless cores to which grains are aligned with the magnetic field remains uncertain. Magnetic fields in protostellar cores are more easily detectable, and generally seen to be ordered and, on the scales observable with single-dish instrumentation, aligned either parallel or perpendicular to outflow direction.

There remain many unanswered questions about the three-dimensional geometry, strength, dynamic importance, and physical role of magnetic fields in star-forming regions. The current and forthcoming generation of submillimeter polarimeters will expand significantly our measurements of magnetic fields; this, coupled with detailed comparison to models, should allow these questions to be addressed.

## AUTHOR CONTRIBUTIONS

## FUNDING

KP acknowledges support from the Ministry of Science and Technology, Taiwan (Grant No. 106-2119-M-007-021-MY3). LF is a Jansky Fellow of the National Radio Astronomy Observatory (NRAO). NRAO is a facility of the National Science Foundation (NSF, operated under cooperative agreement by Associated Universities, Inc.).

## REFERENCES

Adam, R., Adane, A., Ade, P. A. R., André, P., Andrianasolo, A., Aussel, H., et al. (2018). The NIKA2 large-field-of-view millimetre continuum camera for the 30 m IRAM telescope. *Astron. Astrophys.* 609:A115. doi: 10.1051/0004-6361/201731503

Alves, F. O., Frau, P., Girart, J. M., Franco, G. A. P., Santos, F. P., and Wiesemeyer, H. (2014). On the radiation driven alignment of dust grains: detection of the polarization hole in a starless core. *Astron. Astrophys.* 569:L1. doi: 10.1051/0004-6361/201424678

Alves, F. O., Frau, P., Girart, J. M., Franco, G. A. P., Santos, F. P., and Wiesemeyer, H. (2015). On the radiation driven alignment of dust grains: detection of the polarization hole in a starless core (Corrigendum). *Astron. Astrophys.* 574:C4. doi: 10.1051/0004-6361/201424678e

Alves, J. F., Lada, C. J., and Lada, E. A. (2001). Internal structure of a cold dark molecular cloud inferred from the extinction of background starlight. *Nature* 409, 159–161. doi: 10.1038/35051509

Andersson, B.-G., Lazarian, A., and Vaillancourt, J. E. (2015). Interstellar dust grain alignment. *Annu. Rev. Astron. Astrophys.* 53, 501–539. doi: 10.1146/annurev-astro-082214-122414

André, P., Di Francesco, J., Ward-Thompson, D., Inutsuka, S.-I., Pudritz, R. E., and Pineda, J. E. (2014). "From filamentary networks to dense cores in molecular clouds: toward a new paradigm for star formation," in *Protostars and Planets VI*, eds B. Klessen and D. Henning (Tucson: University of Arizona Press), 27–51.

Andre, P., Ward-Thompson, D., and Barsony, M. (1993). Submillimeter continuum observations of Rho Ophiuchi A - The candidate protostar VLA 1623 and prestellar clumps. *Astrophys. J.* 406, 122–141. doi: 10.1086/172425

Attard, M., Houde, M., Novak, G., Li, H.-B., Vaillancourt, J. E., Dowell, C. D., et al. (2009). Magnetic fields and infall motions in NGC 1333 IRAS 4. *Astrophys. J.* 702, 1584–1592. doi: 10.1088/0004-637X/702/2/1584

Austermann, J. E., Beall, J. A., Bryan, S. A., Dober, B., Gao, J., Hilton, G., et al. (2018). Millimeter-wave polarimeters using kinetic inductance detectors for TolTEC and beyond. *J. Low Temp. Phys.* 193, 120–127. doi: 10.1007/s10909-018-1949-5

Bally, J. (2008). "Overview of the Orion complex," in *Handbook of Star Forming Regions, Volume I*, ASP Monograph Series. ed B. Reipurth (San Francisco, CA: Astronomical Society of the Pacific Monograph Publications), 459.

Barvainis, R., Clemens, D. P., and Leach, R. (1988). Polarimetry at 1.3 MM using MILLIPOL - Methods and preliminary results for Orion. *Astron. J.* 95, 510–515. doi: 10.1086/114650

Bastien, P., Bissonnette, E., Simon, A., Coudé, S., Ade, P., Savini, G., et al. (2011). "POL-2: the SCUBA-2 polarimeter," in *Astronomical Polarimetry 2008: Science from Small to Large Telescopes*, volume 449 of *Astronomical Society of the Pacific Conference Series*, eds P. Bastien, N. Manset, D. P. Clemens, and N. St-Louis (San Francisco, CA: Astronomical Society of the Pacific Conference Series), 68.

Bastien, P., Jenness, T., and Molnar, J. (2005). "A polarimeter for SCUBA-2," in *Astronomical Polarimetry: Current Status and Future Directions*, volume 343 of *Astronomical Society of the Pacific Conference Series*, eds A. Adamson, C.

Aspin, C. Davis, and T. Fujiyoshi (San Francisco, CA: Astronomical Society of the Pacific Conference Series), 69.

Basu, S. (2000). Magnetic fields and the triaxiality of molecular cloud cores. *Astrophys. J. Lett.* 540, L103–L106. doi: 10.1086/312885

Benson, P. J., and Myers, P. C. (1989). A survey for dense cores in dark clouds. *Astrophys. J. Suppl.* 71, 89–108. doi: 10.1086/191365

Bok, B. J., and Reilly, E. F. (1947). Small dark nebulae. *Astrophys. J.* 105:255. doi: 10.1086/144901

Bonnor, W. B. (1956). Boyle's Law and gravitational instability. *Month. Notices RAS* 116:351. doi: 10.1093/mnras/116.3.351

Boss, A. P., and Keiser, S. A. (2013). Collapse and fragmentation of magnetic molecular cloud cores with the Enzo AMR MHD Code. I. Uniform density spheres. *Astrophys. J.* 764:136. doi: 10.1088/0004-637X/764/2/136

Brauer, R., Wolf, S., and Reissl, S. (2016). On the origins of polarization holes in Bok globules. *Astron. Astrophys.* 588:A129. doi: 10.1051/0004-6361/201527546

Bryan, S., Austermann, J., Ferrusca, D., Mauskopf, P., McMahon, J., Montana, A., et al. (2018). Optical design of the TolTEC millimeter-wave camera. *ArXiv e-prints*. doi: 10.1117/12.2314130

Burge, C. A., Van Loo, S., Falle, S. A. E. G., and Hartquist, T. W. (2016). Ambipolar diffusion regulated collapse of filaments threaded by perpendicular magnetic fields. *Astron. Astrophys.* 596:A28. doi: 10.1051/0004-6361/201629039

Cashman, L. R., and Clemens, D. P. (2014). The magnetic field of cloud 3 in L204. *Astrophys. J.* 793:126. doi: 10.1088/0004-637X/793/2/126

Chandrasekhar, S., and Fermi, E. (1953). Magnetic fields in spiral arms. *Astrophys. J.* 118:113. doi: 10.1086/145731

Chapman, N. L., Davidson, J. A., Goldsmith, P. F., Houde, M., Kwon, W., Li, Z.-Y., et al. (2013). Alignment between flattened protostellar infall envelopes and ambient magnetic fields. *Astrophys. J.* 770:151. doi: 10.1088/0004-637X/770/2/151

Chapman, N. L., Goldsmith, P. F., Pineda, J. L., Clemens, D. P., Li, D., and Krčo, M. (2011). The magnetic field in Taurus probed by infrared polarization. *Astrophys. J.* 741:21. doi: 10.1088/0004-637X/741/1/21

Chen, C.-Y., King, P. K., and Li, Z.-Y. (2016). Change of magnetic field-gas alignment at the gravity-driven Alfvénic transition in molecular clouds: implications for dust polarization observations. *Astrophys. J.* 829:84. doi: 10.3847/0004-637X/829/2/84

Ching, T.-C., Lai, S.-P., Zhang, Q., Girart, J. M., Qiu, K., and Liu, H. B. (2017). Magnetic fields in the massive dense cores of the DR21 filament: weakly magnetized cores in a strongly magnetized filament. *Astrophys. J.* 838:121. doi: 10.3847/1538-4357/aa65cc

Cho, J., and Yoo, H. (2016). A technique for constraining the driving scale of turbulence and a modified chandrasekhar-fermi method. *Astrophys. J.* 821:21. doi: 10.3847/0004-637X/821/1/21

Chrysostomou, A., Aitken, D. K., Jenness, T., Davis, C. J., Hough, J. H., Curran, R., et al. (2002). The magnetic field structure in W51A. *Astron. Astrophys.* 385, 1014–1021. doi: 10.1051/0004-6361:20020154

Ciolek, G. E., and Basu, S. (2000). Consistency of ambipolar diffusion models with infall in the L1544 protostellar core. *Astrophys. J.* 529, 925–931. doi: 10.1086/308293

Ciolek, G. E., and Mouschovias, T. C. (1994). Ambipolar diffusion, interstellar dust, and the formation of cloud cores and protostars. 3: typical axisymmetric solutions. *Astrophys. J.* 425, 142–160. doi: 10.1086/173971

Clemens, D. P., Leach, R. W., Barvainis, R., and Kane, B. D. (1990). Millipol, a millimeter/submillimeter wavelength polarimeter - Instrument, operation, and calibration. *Publ. ASP* 102, 1064–1076. doi: 10.1086/132735

Coppin, K. E. K., Greaves, J. S., Jenness, T., and Holland, W. S. (2000). Structure, star formation and magnetic fields in the OMC1 region. *Astron. Astrophys.* 356, 1031–1038. Available online at: http://aa.springer.de/bibs/0356003/2301031/small.htm

Cox, N. L. J., Arzoumanian, D., André, P., Rygl, K. L. J., Prusti, T., Men'shchikov, A., et al. (2016). Filamentary structure and magnetic field orientation in Musca. *Astron. Astrophys.* 590:A110. doi: 10.1051/0004-6361/201527068

Crutcher, R. M. (2012). Magnetic fields in molecular clouds. *Annu. Rev. Astron. Astrophys.* 50, 29–63. doi: 10.1146/annurev-astro-081811-125514

Crutcher, R. M., Nutter, D. J., Ward-Thompson, D., and Kirk, J. M. (2004). SCUBA polarization measurements of the magnetic field strengths in the L183, L1544, and L43 prestellar cores. *Astrophys. J.* 600, 279–285. doi: 10.1086/379705

Crutcher, R. M., Troland, T. H., Lazareff, B., and Kazes, I. (1996). CN zeeman observations of molecular cloud cores. *Astrophys. J.* 456:217. doi: 10.1086/176642

Crutcher, R. M., Wandelt, B., Heiles, C., Falgarone, E., and Troland, T. H. (2010). Magnetic fields in interstellar clouds from Zeeman observations: inference of total field strengths by Bayesian analysis. *Astrophys. J.* 725, 466–479. doi: 10.1088/0004-637X/725/1/466

Cudlip, W., Furniss, I., King, K. J., and Jennings, R. E. (1982). Far infrared polarimetry of W51A and M42. *Month. Notices RAS* 200, 1169–1173. doi: 10.1093/mnras/200.4.1169

Curran, R. L., and Chrysostomou, A. (2007). Magnetic fields in massive star-forming regions. *Month. Notices RAS* 382, 699–716. doi: 10.1111/j.1365-2966.2007.12399.x

Curran, R. L., Chrysostomou, A., Collett, J. L., Jenness, T., and Aitken, D. K. (2004). First polarimetry results of two candidate high-mass protostellar objects. *Astron. Astrophys.* 421, 195–202. doi: 10.1051/0004-6361:20034481

Davidson, J. A., Novak, G., Matthews, T. G., Matthews, B., Goldsmith, P. F., Chapman, N., et al. (2011). Magnetic field structure around low-mass Class 0 protostars: B335, L1527, and IC348-SMM2. *Astrophys. J.* 732:97. doi: 10.1088/0004-637X/732/2/97

Davis, C. J., Chrysostomou, A., Matthews, H. E., Jenness, T., and Ray, T. P. (2000). Submillimeter polarimetry of the protostellar outflow sources in Serpens with the Submillimeter Common-User Bolometer Array. *Astrophys. J. Lett.* 530, L115–L118. doi: 10.1086/312476

Davis, L. (1951). The strength of interstellar magnetic fields. *Phys. Rev.* 81, 890–891. doi: 10.1103/PhysRev.81.890.2

Davis, L. Jr., and Greenstein, J. L. (1951). The polarization of starlight by aligned dust grains. *Astrophys. J.* 114:206. doi: 10.1086/145464

Dib, S., Kim, J., Vázquez-Semadeni, E., Burkert, A., and Shadmehri, M. (2007). The virial balance of clumps and cores in molecular clouds. *Astrophys. J.* 661, 262–284. doi: 10.1086/513708

Dolginov, A. Z., and Mitrofanov, I. G. (1976). Orientation of cosmic dust grains. *Astrophys. Space Sci.* 43, 291–317. doi: 10.1007/BF00640010

Dotson, J. L., Davidson, J., Dowell, C. D., Schleuning, D. A., and Hildebrand, R. H. (2000). Far-infrared polarimetry of galactic clouds from the Kuiper Airborne Observatory. *Astrophys. J. Suppl.* 128, 335–370. doi: 10.1086/313384

Dotson, J. L., Novak, G., Renbarger, T., Pernic, D., and Sundwall, J. L. (1998). "SPARO: the submillimeter polarimeter for Antarctic remote observing," in *Advanced Technology MMW, Radio, and Terahertz Telescopes*, volume 3357 of *Proc. SPIE*, ed T. G. Phillips (Bellingham: SPIE (The International Society for Optics and Photonics)), 543–547.

Dotson, J. L., Vaillancourt, J. E., Kirby, L., Dowell, C. D., Hildebrand, R. H., and Davidson, J. A. (2010). 350 $\mu$m polarimetry from the Caltech Submillimeter Observatory. *Astrophys. J. Suppl.* 186, 406–426. doi: 10.1088/0067-0049/186/2/406

Dowell, C. D., HAWC+ Instrument Team, and HAWC+ Science Team (2018). "First Results on Interstellar Magnetic fields from the HAWC+ Instrument for SOFIA," *Americal Astronomical Society (AAS) Meeting Abstracts* Vol. 232, 103.05. Available online at: http://adsabs.harvard.edu/abs/2018AAS...23210305D

Dowell, C. D., Hildebrand, R. H., Schleuning, D. A., Vaillancourt, J. E., Dotson, J. L., Novak, G., et al. (1998). Submillimeter array polarimetry with Hertz. *Astrophys. J.* 504, 588–598. doi: 10.1086/306069

Dragovan, M. (1986). Submillimeter polarization in the Orion Nebula. *Astrophys. J.* 308, 270–280. doi: 10.1086/164498

Draine, B. T., and Weingartner, J. C. (1996). Radiative torques on interstellar grains. I. Superthermal spin-up. *Astrophys. J.* 470:551. doi: 10.1086/177887

Ebert, R. (1955). Über die Verdichtung von H I-Gebieten. Mit 5 Textabbildungen. *Zeitschrift Astrophysik* 37:217.

Elias, J. H. (1978). An infrared study of the Ophiuchus dark cloud. *Astrophys. J.* 224:453. doi: 10.1086/156393

Falceta-Gonçalves, D., Lazarian, A., and Kowal, G. (2008). Studies of regular and random magnetic fields in the ISM: statistics of polarization vectors and the Chandrasekhar-Fermi technique. *Astrophys. J.* 679, 537–551. doi: 10.1086/587479

Fiege, J. D., and Pudritz, R. E. (2000). Helical fields and filamentary molecular clouds - I. *Month. Notices RAS* 311, 85–104. doi: 10.1046/j.1365-8711.2000.03066.x

Fissel, L. M., Ade, P. A. R., Angilè, F. E., Ashton, P., Benton, S. J., Chen, C.-Y., et al. (2018). Relative alignment between the magnetic field and molecular gas structure in the Vela C giant molecular cloud using low and high density tracers. *ArXiv e-prints*. doi: 10.1088/0004-637X/797/1/27

Fissel, L. M., Ade, P. A. R., Angilè, F. E., Ashton, P., Benton, S. J., Devlin, M. J., et al. (2016). Balloon-borne submillimeter polarimetry of the Vela C molecular cloud: systematic dependence of polarization fraction on column density and local polarization-angle dispersion. *Astrophys. J.* 824:134. doi: 10.3847/0004-637X/824/2/134

Flett, A. M., and Murray, A. G. (1991). First results from a submillimetre polarimeter on the James Clerk Maxwell Telescope. *Month. Notices RAS* 249, 4P–6P. doi: 10.1093/mnras/249.1.4P

Foënard, G., Mangilli, A., Aumont, J., Hughes, A., Mot, B., Bernard, J., et al. (2018). PILOT balloon-borne experiment in-flight performance. *ArXiv e-prints*. Available online at: https://arxiv.org/abs/1804.05645

Franzmann, E. L., and Fiege, J. D. (2017). PolCat: modelling submillimetre polarization of molecular cloud cores using successive parametrized coordinate transformations. *Month. Notices RAS* 466, 4592–4613.

Friberg, P., Bastien, P., Berry, D., Savini, G., Graves, S. F., and Pattle, K. (2016). "POL-2: a polarimeter for the James-Clerk-Maxwell telescope," in *Millimeter, Submillimeter, and Far-Infrared Detectors and Instrumentation for Astronomy VIII*, volume 9914 of *Proc. SPIE*, ed H. Zmuidzinas (Bellingham: SPIE (The International Society for Optics and Photonics)), 991403.

Friesen, R. K., Di Francesco, J., Bourke, T. L., Caselli, P., Jørgensen, J. K., Pineda, J. E., et al. (2014). Revealing $H_2D^+$ depletion and compact structure in starless and protostellar cores with ALMA. *Astrophys. J.* 797:27.

Galitzki, N., Ade, P. A. R., Angilè, F. E., Ashton, P., Beall, J. A., Becker, D., et al. (2014a) The next generation BLAST experiment. *J. Astron. Instrument.* 3:1440001. doi: 10.1142/S2251171714400017

Galitzki, N., Ade, P. A. R., Angilè, F. E., Benton, S. J., Devlin, M. J., Dober, B., et al. (2014b). "The balloon-borne large aperture submillimeter telescope for polarimetry-BLASTPol: performance and results from the 2012 Antarctic flight," in *Ground-based and Airborne Telescopes V, volume 9145 of Proc. SPIE*, eds L. M. Stepp, R. Gilmozzi, and H. J. Hall (Bellingham: SPIE (The International Society for Optics and Photonics)), 91450R.

Gaspar Venancio, L. M., Doyle, D., Isaak, K., Onaka, T., Kaneda, H., Nakagawa, T., et al. (2017). "The SPICA telescope: design evolution and expected performance," in *Society of Photo-Optical Instrumentation Engineers (SPIE) Conference Series*, volume 10565 of *Society of Photo-Optical Instrumentation Engineers (SPIE) Conference Series* (Bellingham), 1056555.

Girart, J. M., Rao, R., and Marrone, D. P. (2006). Magnetic fields in the formation of sun-like stars. *Science* 313, 812–814. doi: 10.1126/science.1129093

Goldsmith, P. F., Heyer, M., Narayanan, G., Snell, R., Li, D., and Brunt, C. (2008). Large-scale structure of the molecular gas in Taurus revealed by high linear dynamic range spectral line mapping. *Astrophys. J.* 680, 428–445. doi: 10.1086/587166

González-Casanova, D. F., and Lazarian, A. (2017). Velocity gradients as a tracer for magnetic fields. *Astrophys. J.* 835:41. doi: 10.3847/1538-4357/835/1/41

Greaves, J. S., Holland, W. S., Jenness, T., Chrysostomou, A., Berry, D. S., Murray, A. G., et al. (2003). A submillimetre imaging polarimeter at the James Clerk Maxwell Telescope. *Month. Notices RAS* 340, 353–361. doi: 10.1046/j.1365-8711.2003.06230.x

Greaves, J. S., Holland, W. S., and Ward-Thompson, D. (1997). Submillimeter polarimetry of class 0 protostars: constraints on magnetized outflow models. *Astrophys. J.* 480, 255–261. doi: 10.1086/303970

Güsten, R., Nyman, L. Å., Schilke, P., Menten, K., Cesarsky, C., and Booth, R. (2006). The Atacama Pathfinder EXperiment (APEX) - a new submillimeter facility for southern skies -. *Astron. Astrophys.* 454, L13–L16. doi: 10.1051/0004-6361:20065420

Harper, D. A., Runyan, M. C., Dowell, C. D., Wirth, C. J., Amato, M., Ames, T., et al. (2018). HAWC+, the far-infrared camera and polarimeter for SOFIA. *J. Astron. Instrum.* 7, 1840008–1025. doi: 10.1142/S2251171718400081

Heiles, C. (1996). The local direction and curvature of the galactic magnetic field derived from starlight polarization. *Astrophys. J.* 462:316. doi: 10.1086/177153

Heiles, C., and Robishaw, T. (2009). "Zeeman splitting in the diffuse interstellar medium-The Milky Way and beyond," in *Cosmic Magnetic Fields: From Planets, to Stars and Galaxies*, volume 259 of *IAU Symposium*, eds K. G. Strassmeier, A. G. Kosovichev, and J. E. Beckman (Cambridge, UK), 579–590.

Heitsch, F., Zweibel, E. G., Mac Low, M.-M., Li, P., and Norman, M. L. (2001). Magnetic field diagnostics based on far-infrared polarimetry: tests using numerical simulations. *Astrophys. J.* 561, 800–814. doi: 10.1086/323489

Henning, T., Wolf, S., Launhardt, R., and Waters, R. (2001). Measurements of the magnetic field geometry and strength in Bok globules. *Astrophys. J.* 561, 871–879. doi: 10.1086/323362

Hester, J. J., Scowen, P. A., Sankrit, R., Lauer, T. R., Ajhar, E. A., Baum, W. A., et al. (1996). Hubble space telescope WFPC2 imaging of M16: photoevaporation and emerging young stellar objects. *Astron. J.* 111:2349. doi: 10.1086/117968

Hildebrand, R. H., Davidson, J. A., Dotson, J. L., Dowell, C. D., Novak, G., and Vaillancourt, J. E. (2000). A primer on far-infrared polarimetry. *Publ. ASP* 112, 1215–1235. doi: 10.1086/316613

Hildebrand, R. H., Dragovan, M., and Novak, G. (1984). Detection of submillimeter polarization in the Orion nebula. *Astrophys. J. Lett.* 284, L51–L54. doi: 10.1086/184351

Hildebrand, R. H., Kirby, L., Dotson, J. L., Houde, M., and Vaillancourt, J. E. (2009). Dispersion of magnetic fields in molecular clouds. I. *Astrophys. J.* 696, 567–573. doi: 10.1088/0004-637X/696/1/567

Holland, W. S., Greaves, J. S., Ward-Thompson, D., and Andre, P. (1996). The magnetic field structure around protostars. Submillimetre polarimetry of VLA 1623 and S 106-IR/FIR. *Astron. Astrophys.* 309, 267–274.

Houde, M., Dowell, C. D., Hildebrand, R. H., Dotson, J. L., Vaillancourt, J. E., Phillips, T. G., et al. (2004). Tracing the magnetic field in Orion A. *Astrophys. J.* 604, 717–740. doi: 10.1086/382067

Houde, M., Vaillancourt, J. E., Hildebrand, R. H., Chitsazzadeh, S., and Kirby, L. (2009). Dispersion of magnetic fields in molecular clouds. II. *Astrophys. J.* 706, 1504–1516. doi: 10.1088/0004-637X/706/2/1504

Hull, C. L. H., Girart, J. M., Tychoniec, Ł., Rao, R., Cortés, P. C., Pokhrel, R., et al. (2017). ALMA observations of dust polarization and molecular line emission from the Class 0 protostellar source Serpens SMM1. *Astrophys. J.* 847:92. doi: 10.3847/1538-4357/aa7fe9

Hull, C. L. H., Plambeck, R. L., Kwon, W., Bower, G. C., Carpenter, J. M., Crutcher, R. M., et al. (2014). TADPOL: A 1.3 mm survey of dust polarization in star-forming cores and regions. *Astrophys. J. Suppl.* 213:13. doi: 10.1088/0067-0049/213/1/13

Hull, C. L. H., and Zhang, Q. (2019). Interferometric observations of magnetic fields in forming stars. *Front. Astron. Space Sci.* 6:3. doi: 10.3389/fspas.2019.00003

Jones, T. J., Bagley, M., Krejny, M., Andersson, B.-G., and Bastien, P. (2015). Grain alignment in starless cores. *Astron. J.* 149:31. doi: 10.1088/0004-6256/149/1/31

Jones, T. J., Gordon, M., Shenoy, D., Gehrz, R. D., Vaillancourt, J. E., and Krejny, M. (2016). SOFIA mid-infrared imaging and CSO submillimeter polarimetry observations of G034.43+00.24 MM1. *Astron. J.* 151:156. doi: 10.3847/0004-6256/151/6/156

Jow, D. L., Hill, R., Scott, D., Soler, J. D., Martin, P. G., Devlin, M. J., et al. (2018). An application of an optimal statistic for characterizing relative orientations. *Mon. Not. R. Astron. Soc.* 474, 1018–1027. doi: 10.1093/mnras/stx2736

Kirk, J. M., Ward-Thompson, D., and Crutcher, R. M. (2006). SCUBA polarization observations of the magnetic fields in the pre-stellar cores L1498 and L1517B. *Month. Notices RAS* 369, 1445–1450. doi: 10.1111/j.1365-2966.2006.10392.x

Klassen, M., Pudritz, R. E., and Kirk, H. (2017). Filamentary flow and magnetic geometry in evolving cluster-forming molecular cloud clumps. *Month. Notices RAS* 465, 2254–2276. doi: 10.1093/mnras/stw2889

Koch, P. M., Tang, Y.-W., and Ho, P. T. P. (2012a). Magnetic field strength maps for molecular clouds: a new method based on a polarization-intensity gradient relation. *Astrophys. J.* 747:79. doi: 10.1088/0004-637X/747/1/79

Koch, P. M., Tang, Y.-W., and Ho, P. T. P. (2012b). Quantifying the significance of the magnetic field from large-scale cloud to collapsing core: self-similarity, mass-to-flux ratio, and star formation efficiency. *Astrophys. J.* 747:80. doi: 10.1088/0004-637X/747/1/80

Koch, P. M., Tang, Y.-W., and Ho, P. T. P. (2013). Interpreting the role of the magnetic field from dust polarization maps. *Astrophys. J.* 775:77. doi: 10.1088/0004-637X/775/1/77

Krejny, M., Li, H., Matthews, T., Novak, G., Shinnaga, H., Vaillancourt, J., et al. (2011). "Submillimeter spectropolarimetry as a probe for grain growth in DG Tau," in *Astronomical Polarimetry 2008: Science from Small to Large Telescopes*, volume 449 of *Astronomical Society of the Pacific Conference Series*, eds P. Bastien, N. Manset, D. P. Clemens, and N. St-Louis (San Francisco, CA), 338.

Kusune, T., Sugitani, K., Nakamura, F., Watanabe, M., Tamura, M., Kwon, J., et al. (2016). Magnetic field of the Vela C molecular cloud. *Astrophys. J. Lett.* 830:L23. doi: 10.3847/2041-8205/830/2/L23

Kwon, J., Doi, Y., Tamura, M., Matsumura, M., Pattle, K., Berry, D., et al. (2018). A first look at BISTRO observations of the $\rho$ Oph-A core. *Astrophys. J.* 859:4. doi: 10.3847/1538-4357/aabd82

Lada, C. J. (1987). "Star formation - From OB associations to protostars," in *Star Forming Regions, volume 115 of IAU Symposium*, eds M. Peimbert and J. Jugaku (Dordrecht), 1–17.

Lai, S.-P., Crutcher, R. M., Girart, J. M., and Rao, R. (2002). Interferometric mapping of magnetic fields in star-forming regions. II. NGC 2024 FIR 5. *Astrophys. J.* 566, 925–930. doi: 10.1086/338336

Lamarre, J.-M., Puget, J.-L., Ade, P. A. R., Bouchet, F., Guyot, G., Lange, A. E., et al. (2010). Planck pre-launch status: the HFI instrument, from specification to actual performance. *Astron. Astrophys.* 520:A9. doi: 10.1051/0004-6361/200912975

Launhardt, R., Nutter, D., Ward-Thompson, D., Bourke, T. L., Henning, T., Khanzadyan, T., et al. (2010). Looking into the hearts of Bok globules: millimeter and submillimeter continuum images of isolated star-forming cores. *Astrophys. J. Suppl.* 188, 139–177. doi: 10.1088/0067-0049/188/1/139

Lazarian, A., and Hoang, T. (2007). Radiative torques: analytical model and basic properties. *Month. Notices RAS* 378, 910–946. doi: 10.1111/j.1365-2966.2007.11817.x

Levin, S. M., Langer, W. D., Velusamy, T., Kuiper, T. B. H., and Crutcher, R. M. (2001). Measuring the magnetic field strength in L1498 with Zeeman-splitting observations of CCS. *Astrophys. J.* 555, 850–854. doi: 10.1086/321518

Li, H., Dowell, C. D., Kirby, L., Novak, G., and Vaillancourt, J. E. (2008). Design and initial performance of SHARP, a polarimeter for the SHARC-II camera at the Caltech Submillimeter Observatory. *Appl. Opt.* 47, 422–430. doi: 10.1364/AO.47.000422

Li, H., Griffin, G. S., Krejny, M., Novak, G., Loewenstein, R. F., Newcomb, M. G., et al. (2006). Results of SPARO 2003: mapping magnetic fields in giant molecular clouds. *Astrophys. J.* 648, 340–354. doi: 10.1086/505858

Li, H.-B., Dowell, C. D., Goodman, A., Hildebrand, R., and Novak, G. (2009). Anchoring magnetic field in turbulent molecular clouds. *Astrophys. J.* 704, 891–897. doi: 10.1088/0004-637X/704/2/891

Li, H.-B., and Henning, T. (2011). The alignment of molecular cloud magnetic fields with the spiral arms in M33. *Nature* 479, 499–501. doi: 10.1038/nature10551

Li, P. S., McKee, C. F., and Klein, R. I. (2015). Magnetized interstellar molecular clouds - I. Comparison between simulations and Zeeman observations. *Month. Notices RAS* 452, 2500–2527. doi: 10.1093/mnras/stv1437

Liu, T., Li, P. S., Juvela, M., Kim, K.-T., Evans, II, N. J., Di Francesco, J., et al. (2018). A holistic perspective on the dynamics of G035.39-00.33: the interplay between gas and magnetic fields. *Astrophys. J.* 859:151. doi: 10.3847/1538-4357/aac025

Machida, M. N., Tomisaka, K., and Matsumoto, T. (2004). First MHD simulation of collapse and fragmentation of magnetized molecular cloud cores. *Month. Notices RAS* 348, L1–L5. doi: 10.1111/j.1365-2966.2004.07402.x

Marsden, G., Ade, P. A. R., Bock, J. J., Chapin, E. L., Devlin, M. J., Dicker, S. R., et al. (2009). BLAST: resolving the cosmic submillimeter background. *Astrophys. J.* 707, 1729–1739. doi: 10.1088/0004-637X/707/2/1729

Matthews, B. C., Fiege, J. D., and Moriarty-Schieven, G. (2002). Magnetic fields in star-forming molecular clouds. III. Submillimeter polarimetry of intermediate-mass cores and filaments in Orion B. *Astrophys. J.* 569, 304–321. doi: 10.1086/339318

Matthews, B. C., Lai, S.-P., Crutcher, R. M., and Wilson, C. D. (2005). Multiscale magnetic fields in star-forming regions: interferometric polarimetry of the MMS 6 core of OMC-3. *Astrophys. J.* 626, 959–965. doi: 10.1086/430127

Matthews, B. C., McPhee, C. A., Fissel, L. M., and Curran, R. L. (2009). The legacy of SCUPOL: 850 $\mu$m imaging polarimetry from 1997 to 2005. *Astrophys. J. Suppl.* 182, 143–204. doi: 10.1088/0067-0049/182/1/143

Matthews, B. C., and Wilson, C. D. (2002a). Magnetic fields in star-forming molecular clouds. IV. Polarimetry of the filamentary NGC 2068 cloud in Orion B. *Astrophys. J.* 571, 356–365. doi: 10.1086/339915

Matthews, B. C., and Wilson, C. D. (2002b). Magnetic fields in star-forming molecular clouds. V. Submillimeter polarization of the Barnard 1 dark cloud. *Astrophys. J.* 574, 822–833. doi: 10.1086/341111

Matthews, B. C., Wilson, C. D., and Fiege, J. D. (2001a). "Magnetic fields in star-forming clouds: how can FIRST contribute?," in *The Promise of the Herschel Space Observatory*, volume 460 of *ESA Special Publication*, eds G. L. Pilbratt, J. Cernicharo, A. M. Heras, T. Prusti, and R. Harris (Noordwijk: ESA Publications), 463.

Matthews, B. C., Wilson, C. D., and Fiege, J. D. (2001b). Magnetic fields in star-forming molecular clouds. II. The depolarization effect in the OMC-3 filament of Orion A. *Astrophys. J.* 562, 400–423. doi: 10.1086/323375

Matthews, T. G., Ade, P. A. R., Angilè, F. E., Benton, S. J., Chapin, E. L., Chapman, N. L., et al. (2014). Lupus I observations from the 2010 flight of the balloon-borne large aperture submillimeter telescope for polarimetry. *Astrophys. J.* 784:116. doi: 10.1088/0004-637X/784/2/116

Mauskopf, P. D. (2018). Transition edge sensors and kinetic inductance detectors in astronomical instruments. *Publ. ASP* 130:082001.

Minchin, N. R., Sandell, G., and Murray, A. G. (1995). Submm polarimetric observations of NGC 1333 IRAS 4A and 4B: tracing the circumstellar magnetic field. *Astron. Astrophys.* 293, L61–L64.

Mocz, P., Burkhart, B., Hernquist, L., McKee, C. F., and Springel, V. (2017). Moving-mesh Simulations of Star-forming Cores in Magneto-gravo-turbulence. *Astrophys. J.* 838:40. doi: 10.3847/1538-4357/aa6475

Monsch, K., Pineda, J. E., Liu, H. B., Zucker, C., How-Huan Chen, H., Pattle, K., et al. (2018). Dense gas kinematics and a narrow filament in the Orion A OMC1 region using $NH_3$. *Astrophys. J.* 861:77. doi: 10.3847/1538-4357/aac8da

Motte, F., André, P., and Neri, R. (1998). The initial conditions of star formation in the rho Ophiuchi main cloud: wide-field millimeter continuum mapping. *Astron. Astrophys.* 336, 150–172.

Mouschovias, T. C. (1976a). Nonhomologous contraction and equilibria of self-gravitating, magnetic interstellar clouds embedded in an intercloud medium: star formation. I Formulation of the problem and method of solution. *Astrophys. J.* 206, 753–767. doi: 10.1086/154436

Mouschovias, T. C. (1976b). Nonhomologous contraction and equilibria of self-gravitating, magnetic interstellar clouds embedded in an intercloud medium: star formation. II - Results. *Astrophys. J.* 207, 141–158. doi: 10.1086/154478

Murray, A. G., Nartallo, R., Haynes, C. V., Gannaway, F., and Ade, P. A. R. (1997). "An imaging polarimeter for SCUBA," in *The Far Infrared and Submillimetre Universe.*, volume 401 of *ESA Special Publication*, ed A. Wilson (Noordwijk: ESA Publications), 405.

Myers, P. C., and Goodman, A. A. (1991). On the dispersion in direction of interstellar polarization. *Astrophys. J.* 373, 509–524. doi: 10.1086/170070

Nakamura, F., Hanawa, T., and Nakano, T. (1993). Fragmentation of filamentary molecular clouds with longitudinal and helical magnetic fields. *Publ. Astron. Soc. Japan* 45, 551–566.

Nakamura, F., and Li, Z.-Y. (2008). Magnetically regulated star formation in three dimensions: the case of the Taurus molecular cloud complex. *Astrophys. J.* 687, 354–375. doi: 10.1086/591641

Novak, G., Dotson, J. L., and Li, H. (2009). Dispersion of observed position angles of submillimeter polarization in molecular clouds. *Astrophys. J.* 695, 1362–1369. doi: 10.1088/0004-637X/695/2/1362

Novak, G., Gonatas, D. P., Hildebrand, R. H., and Platt, S. R. (1989). A 100-micron polarimeter for the Kuiper Airborne Observatory. *Publ. ASP* 101, 215–224. doi: 10.1086/132425

Ossenkopf, V., and Henning, T. (1994). Dust opacities for protostellar cores. *Astron. Astrophys.* 291, 943–959.

Ostriker, E. C., Stone, J. M., and Gammie, C. F. (2001). Density, velocity, and magnetic field structure in turbulent molecular cloud models. *Astrophys. J.* 546, 980–1005. doi: 10.1086/318290

Ostriker, J. (1964). The equilibrium of polytropic and isothermal cylinders. *Astrophys. J.* 140:1056. doi: 10.1086/148005

Otal, L. E. (2014). *The Optical System and the Astronomical Potential of A-MKID, a New Camera Using Microwave Kinetic Inductance Detector Technolog*. PhD thesis, University of Bonn.

Padoan, P., Goodman, A., Draine, B. T., Juvela, M., Nordlund, Å., and Rögnvaldsson, Ö. E. (2001). Theoretical models of polarized dust emission from protostellar cores. *Astrophys. J.* 559, 1005–1018. doi: 10.1086/322504

Padovani, M., Brinch, C., Girart, J. M., Jørgensen, J. K., Frau, P., Hennebelle, P., et al. (2012). Adaptable radiative transfer innovations for submillimetre telescopes (ARTIST). Dust polarisation module (DustPol). *Astron. Astrophys.* 543:A16. doi: 10.1051/0004-6361/201219028

Palmeirim, P., André, P., Kirk, J., Ward-Thompson, D., Arzoumanian, D., Könyves, V., et al. (2013). Herschel view of the Taurus B211/3 filament and striations: evidence of filamentary growth? *Astron. Astrophys.* 550:A38. doi: 10.1051/0004-6361/201220500

Panopoulou, G. V., Psaradaki, I., and Tassis, K. (2016). The magnetic field and dust filaments in the Polaris Flare. *Month. Notices RAS* 462, 1517–1529. doi: 10.1093/mnras/stw1678

Pattle, K., Ward-Thompson, D., Berry, D., Hatchell, J., Chen, H.-R., Pon, A., et al. (2017). The JCMT BISTRO survey: the magnetic field strength in the Orion A filament. *Astrophys. J.* 846:122. doi: 10.3847/1538-4357/aa80e5

Pattle, K., Ward-Thompson, D., Hasegawa, T., Bastien, P., Kwon, W., Lai, S.-P., et al. (2018). First observations of the magnetic field inside the Pillars of Creation: results from the BISTRO survey. *Astrophys. J. Lett.* 860:L6. doi: 10.3847/2041-8213/aac771

Pattle, K., Ward-Thompson, D., Kirk, J. M., White, G. J., Drabek-Maunder, E., Buckle, J., et al. (2015). The JCMT Gould Belt Survey: first results from the SCUBA-2 observations of the Ophiuchus molecular cloud and a virial analysis of its prestellar core population. *Month. Notices RAS* 450, 1094–1122. doi: 10.1093/mnras/stv376

Pillai, T., Kauffmann, J., Tan, J. C., Goldsmith, P. F., Carey, S. J., and Menten, K. M. (2015). Magnetic fields in high-mass infrared dark clouds. *Astrophys. J.* 799:74. doi: 10.1088/0004-637X/799/1/74

Pillai, T., Kauffmann, J., Wiesemeyer, H., and Menten, K. M. (2016). CN Zeeman and dust polarization in a high-mass cold clump. *Astron. Astrophys.* 591:A19. doi: 10.1051/0004-6361/201527803

Planck Collaboration Int. XIX (2015). Planck intermediate results. XIX. An overview of the polarized thermal emission from Galactic dust. *Astron. Astrophys.* 576:A104. doi: 10.1051/0004-6361/201424082

Planck Collaboration Int. XX (2015). Planck intermediate results. XX. Comparison of polarized thermal emission from Galactic dust with simulations of MHD turbulence. *Astron. Astrophys.* 576:A105. doi: 10.1051/0004-6361/201424086

Planck Collaboration Int. XXXIV (2016). Planck intermediate results. XXXIV. The magnetic field structure in the Rosette Nebula. *Astron. Astrophys. A&A* 586:A137. doi: 10.1051/0004-6361/201525616

Planck Collaboration Int. XXXV (2016). Planck intermediate results. XXXV. Probing the role of the magnetic field in the formation of structure in molecular clouds. *Astron. Astrophys. A&A* 586:A138. doi: 10.1051/0004-6361/201525896

Planck Collaboration VIII (2016). Planck 2015 results. VIII. High frequency instrument data processing: calibration and maps. *Astron. Astrophys. A&A* 594:A8. doi: 10.1051/0004-6361/201525820

Planck Collaboration, Ade, P. A. R., Aghanim, N., Alves, M. I. R., Arnaud, M., Arzoumanian, D., et al. (2016). Planck intermediate results. XXXIII. Signature of the magnetic field geometry of interstellar filaments in dust polarization maps. *Astron. Astrophys.* 586:A136. doi: 10.1051/0004-6361/201425305

Planck Collaboration, Aghanim, N., Akrami, Y., Alves, M. I. R., Ashdown, M., Aumont, J., et al. (2018). Planck 2018 results. XII. Galactic astrophysics using polarized dust emission. *ArXiv e-prints*.

Platt, S. R., Hildebrand, R. H., Pernic, R. J., Davidson, J. A., and Novak, G. (1991). 100-micron array polarimetry from the Kuiper Airborne Observatory - Instrumentation, techniques, and first results. *Publ. ASP* 103, 1193–1210.

Plummer, H. C. (1911). On the problem of distribution in globular star clusters. *Month. Notices RAS* 71, 460–470.

Poidevin, F., Bastien, P., and Jones, T. J. (2011). Multi-scale analysis of magnetic fields in filamentary molecular clouds in Orion A. *Astrophys. J.* 741:112. doi: 10.1088/0004-637X/741/2/112

Poidevin, F., Bastien, P., and Matthews, B. C. (2010). Magnetic field structures and turbulent components in the star-forming molecular clouds OMC-2 and OMC-3. *Astrophys. J.* 716, 893–906. doi: 10.1088/0004-637X/716/2/893

Poidevin, F., Falceta-Gonçalves, D., Kowal, G., de Gouveia Dal Pino, E., and Mário Magalhães, A. (2013). Magnetic field components analysis of the SCUPOL 850 µm polarization data catalog. *Astrophys. J.* 777:112. doi: 10.1088/0004-637X/777/2/112

Price, D. J., and Bate, M. R. (2007). The impact of magnetic fields on single and binary star formation. *Month. Notices RAS* 377, 77–90. doi: 10.1111/j.1365-2966.2007.11621.x

Rao, R., Crutcher, R. M., Plambeck, R. L., and Wright, M. C. H. (1998). High-resolution millimeter-wave mapping of linearly polarized dust emission: magnetic field structure in Orion. *Astrophys. J. Lett.* 502, L75–L78. doi: 10.1086/311485

Rathborne, J. M., Jackson, J. M., and Simon, R. (2006). Infrared dark clouds: precursors to star clusters. *Astrophys. J.* 641, 389–405. doi: 10.1086/500423

Reissl, S., Wolf, S., and Brauer, R. (2016). Radiative transfer with POLARIS. I. Analysis of magnetic fields through synthetic dust continuum polarization measurements. *Astron. Astrophys.* 593:A87. doi: 10.1051/0004-6361/201424930

Renbarger, T., Chuss, D. T., Dotson, J. L., Griffin, G. S., Hanna, J. L., Loewenstein, R. F., et al. (2004). Early results from SPARO: instrument characterization and polarimetry of NGC 6334. *Publ. ASP* 116, 415–424. doi: 10.1086/383623

Roelfsema, P. R., Shibai, H., Armus, L., Arrazola, D., Audard, M., Audley, M. D., et al. (2018). SPICA - a large cryogenic infrared space telescope unveiling the obscured Universe. *ArXiv e-prints*. doi: 10.1017/pasa.2018.15

Salji, C. J., Richer, J. S., Buckle, J. V., di Francesco, J., Hatchell, J., Hogerheijde, M., et al. (2015). The JCMT Gould Belt Survey: properties of star-forming filaments in Orion A North. *Month. Notices RAS* 449, 1782–1796. doi: 10.1093/mnras/stv369

Schleicher, D. R. G., and Stutz, A. (2018). Magnetic tension and instabilities in the Orion A integral-shaped filament. *Month. Notices RAS* 475, 121–127. doi: 10.1093/mnras/stx2975

Schleuning, D. A. (1998). Far-infrared and submillimeter polarization of OMC-1: evidence for magnetically regulated star formation. *Astrophys. J.* 493, 811–825. doi: 10.1086/305139

Schleuning, D. A., Dowell, C. D., Hildebrand, R. H., Platt, S. R., and Novak, G. (1997). HERTZ, a submillimeter polarimeter. *Publ. ASP* 109, 307–318. doi: 10.1086/133892

Seifried, D., Walch, S., Reissl, S., and Ibáñez-Mejía, J. C. (2019). SILCC-Zoom: polarisation and depolarisation in molecular clouds. *MNRAS*. 482, 2697–2716. doi: 10.1093/mnras/sty2831

Siringo, G., Kovács, A., Kreysa, E., Schuller, F., Weiss, A., Guesten, R., et al. (2012). "First results of the polarimeter for the Large APEX Bolometer Camera (LABOCA)," in *Millimeter, Submillimeter, and Far-Infrared Detectors and Instrumentation for Astronomy VI, volume 8452 of* Proc. SPIE, eds W. S. Holland and J. Zmuidzinas (Bellingham: SPIE (The International Society for Optics and Photonics)), 845206.

Soam, A., Pattle, K., Ward-Thompson, D., Lee, C. W., Sadavoy, S., Koch, P. M., et al. (2018). Magnetic fields towards Ophiuchus-B derived from SCUBA-2 polarization measurements. *ArXiv e-prints*. doi: 10.3847/1538-4357/aac4a6

Soler, J. D., Ade, P. A. R., Angilè, F. E., Ashton, P., Benton, S. J., Devlin, M. J., et al. (2017). The relation between the column density structures and the magnetic field orientation in the Vela C molecular complex. *Astron. Astrophys.* 603:A64. doi: 10.1051/0004-6361/201730608

Soler, J. D., Bracco, A., and Pon, A. (2018). The magnetic environment of the Orion-Eridanus superbubble as revealed by Planck. *Astron. Astrophys.* 609:L3. doi: 10.1051/0004-6361/201732203

Soler, J. D., and Hennebelle, P. (2017). What are we learning from the relative orientation between density structures and the magnetic field in molecular clouds? *Astron. Astrophys.* 607:A2. doi: 10.1051/0004-6361/201731049

Soler, J. D., Hennebelle, P., Martin, P. G., Miville-Deschênes, M.-A., Netterfield, C. B., and Fissel, L. M. (2013). An imprint of molecular cloud magnetization in the morphology of the dust polarized emission. *Astrophys. J.* 774:128. doi: 10.1088/0004-637X/774/2/128

Staguhn, J., Amatucci, E., Armus, L., Bradley, D., Carter, R., Chuss, D., et al. (2018). Origins space telescope: the far infrared imager and polarimeter FIP. *Proc. SPIE* 10698, 10698 – 10698 – 6. doi: 10.1117/12.2312626

Stephens, I. W., Looney, L. W., Kwon, W., Hull, C. L. H., Plambeck, R. L., Crutcher, R. M., et al. (2013). The magnetic field morphology of the class 0 protostar L1157-mm. *Astrophys. J. Lett.* 769:L15. doi: 10.1088/2041-8205/769/1/L15

Stodółkiewicz, J. S. (1963). On the gravitational instability of some magneto-hydrodynamical systems of astrophysical interest. Part III. *Acta Astron.* 13, 30–54.

Sugitani, K., Nakamura, F., Watanabe, M., Tamura, M., Nishiyama, S., Nagayama, T., et al. (2011). Near-infrared-imaging Polarimetry Toward Serpens South:

revealing the Importance of the Magnetic Field. *Astrophys. J.* 734:63. doi: 10.1088/0004-637X/734/1/63

Sutin, B., Alvarez, M., Battaglia, N., Bock, J., Bonato, M., Borrill, J., et al. (2018). PICO - the probe of inflation and cosmic origins. *ArXiv e-prints*. doi: 10.1117/12.2311326

Tamura, M. (1999). "Submillimeter polarimetry of star forming regions: from cloud cores to circumstellar disks," in *Star Formation 1999*, ed T. Nakamoto (Nobeyama: Nobeyama Radio Observatory), 212–216.

Tamura, M., Hough, J. H., and Hayashi, S. S. (1995). 1 millimeter polarimetry of young stellar objects: low-mass protostars and T Tauri stars. *Astrophys. J.* 448:346. doi: 10.1086/175965

Tan, J. C., Beltrán, M. T., Caselli, P., Fontani, F., Fuente, A., Krumholz, M. R., et al. (2014). "Massive star formation," in *Protostars and Planets VI*, eds H. Beuther, R. Klessen, C. Dullemond, and T. Henning (Tucson: University of Arizona Press), 149–172. doi: 10.2458/azu_uapress_9780816531240-ch007

Tang, Y.-W., Ho, P. T. P., Koch, P. M., Guilloteau, S., and Dutrey, A. (2013). Dust continuum and polarization from envelope to cores in star formation: a case study in the W51 north region. *Astrophys. J.* 763:135. doi: 10.1088/0004-637X/763/2/135

Tassis, K., Dowell, C. D., Hildebrand, R. H., Kirby, L., and Vaillancourt, J. E. (2009). Statistical assessment of shapes and magnetic field orientations in molecular clouds through polarization observations. *Month. Notices RAS* 399, 1681–1693. doi: 10.1111/j.1365-2966.2009.15420.x

Tomisaka, K. (2014). Magnetohydrostatic equilibrium structure and mass of filamentary isothermal cloud threaded by lateral magnetic field. *Astrophys. J.* 785:24. doi: 10.1088/0004-637X/785/1/24

Tomisaka, K. (2015). Polarization structure of filamentary clouds. *Astrophys. J.* 807:47. doi: 10.1088/0004-637X/807/1/47

Valdivia, V., Maury, A., Hennebelle, P., Galametz, M., and Reissl, S. (2017). "Towards realistic predictions of mm/sub-mm polarized dust emission," in *Submm/mm/cm QUESO Workshop 2017 (QUESO2017)* (Garching), 30.

Vallée, J. P., and Bastien, P. (1999). Magnetism in interstellar nurseries at 760 microns. *Astrophys. J.* 526, 819–832. doi: 10.1086/308010

Vallée, J. P., Bastien, P., and Greaves, J. S. (2000). Highly polarized thermal dust emission in the Bok globule CB 068. *Astrophys. J.* 542, 352–358. doi: 10.1086/309531

Vallée, J. P., and Fiege, J. D. (2006). A cool filament crossing the warm protostar DR 21(OH): geometry, kinematics, magnetic vectors, and pressure balance. *Astrophys. J.* 636, 332–347. doi: 10.1086/497957

Vallée, J. P., and Fiege, J. D. (2007). OMC-1: a cool arching filament in a hot gaseous cavity: geometry, kinematics, magnetic vectors, and pressure balance. *Astron. J.* 133, 1012–1026. doi: 10.1086/511004

Vallée, J. P., Greaves, J. S., and Fiege, J. D. (2003). Magnetic structure of a dark Bok globule. *Astrophys. J.* 588, 910–917. doi: 10.1086/374309

Wang, J.-W., Lai, S.-P., Eswaraiah, C., Pattle, K., Di Francesco, J., Johnstone, D., et al. (2018). JCMT BISTRO survey: magnetic fields within the hub-filament structure in IC 5146. *arXiv e-prints*.

Ward-Thompson, D., Kirk, J. M., Crutcher, R. M., Greaves, J. S., Holland, W. S., and André, P. (2000). First observations of the magnetic field geometry in prestellar cores. *Astrophys. J. Lett.* 537, L135–L138. doi: 10.1086/312764

Ward-Thompson, D., Pattle, K., Bastien, P., Furuya, R. S., Kwon, W., Lai, S.-P., et al. (2017). First results from BISTRO – a SCUBA-2 polarimeter survey of the Gould Belt. *Astrophys. J.* 842:66. doi: 10.3847/1538-4357/aa70a0

Ward-Thompson, D., Scott, P. F., Hills, R. E., and Andre, P. (1994). A submillimetre continuum survey of pre protostellar cores. *Month. Notices RAS* 268:276. doi: 10.1093/mnras/268.1.276

Ward-Thompson, D., Sen, A. K., Kirk, J. M., and Nutter, D. (2009). Optical and submillimetre observations of Bok globules - tracing the magnetic field from low to high density. *Month. Notices RAS* 398, 394–400. doi: 10.1111/j.1365-2966.2009.15159.x

Wareing, C. J., Pittard, J. M., Falle, S. A. E. G., and Van Loo, S. (2016). Magnetohydrodynamical simulation of the formation of clumps and filaments in quiescent diffuse medium by thermal instability. *Month. Notices RAS* 459, 1803–1818. doi: 10.1093/mnras/stw581

Whittet, D. C. B., Hough, J. H., Lazarian, A., and Hoang, T. (2008). The efficiency of grain alignment in dense interstellar clouds: a reassessment of constraints from near-infrared polarization. *Astrophys. J.* 674, 304–315. doi: 10.1086/525040

Whitworth, A. P., and Ward-Thompson, D. (2001). An empirical model for protostellar collapse. *Astrophys. J.* 547, 317–322. doi: 10.1086/318373

Wiesemeyer, H., Hezareh, T., Kreysa, E., Weiss, A., Güsten, R., Menten, K. M., et al. (2014). Submillimeter polarimetry with PolKa, a reflection-type modulator for the APEX telescope. *Publ. ASP* 126:1027.

Wolf, S., Launhardt, R., and Henning, T. (2003). Magnetic field evolution in Bok globules. *Astrophys. J.* 592, 233–244. doi: 10.1086/375622

Wu, B., Tan, J. C., Nakamura, F., Van Loo, S., Christie, D., and Collins, D. (2017). GMC collisions as triggers of star formation. II. 3D turbulent, magnetized simulations. *Astrophys. J.* 835:137. doi: 10.3847/1538-4357/835/2/137

Young, K., Alvarez, M., Battaglia, N., Bock, J., Borrill, J., Chuss, D., et al. (2018). Optical design of PICO, a concept for a space mission to probe inflation and cosmic origins. *ArXiv e-prints*. doi: 10.1117/12.2309421

Yuen, K. H., and Lazarian, A. (2017). Tracing interstellar magnetic field using velocity gradient technique: application to atomic hydrogen data. *Astrophys. J. Lett.* 837:L24. doi: 10.3847/2041-8213/aa6255

Zweibel, E. G. (1990). Magnetic field-line tangling and polarization measurements in clumpy molecular gas. *Astrophys. J.* 362, 545–550. doi: 10.1086/169291

**Conflict of Interest Statement:** The authors declare that the research was conducted in the absence of any commercial or financial relationships that could be construed as a potential conflict of interest.

The reviewer TG declared a past co-authorship with one of the authors KP to the handling editor.